\documentclass[nolinenumbers,trackchanges]{aastex701}
\usepackage{multirow}

\begin{document}

\title{The Phenomenological Classification of TESS Eclipsing Binaries}

\author[orcid=0009-0000-8102-3471]{Shi-Qi Liu}
\affiliation{Shandong Key Laboratory of Space Environment and Exploration Technology, Institute of Space Sciences, School of Space Science and Technology, Shandong University, Weihai, Shandong 264209, People's Republic of China}
\email{m15615185800@163.com}

\author[orcid=0000-0003-3590-335X]{Kai Li}
\affiliation{Shandong Key Laboratory of Space Environment and Exploration Technology, Institute of Space Sciences, School of Space Science and Technology, Shandong University, Weihai, Shandong 264209, People's Republic of China}
\email[show]{kaili@sdu.edu.cn}
\correspondingauthor{Kai Li}

\author[orcid=0000-0001-7084-0484]{Xiao-Dian Chen} 
\affiliation{CAS Key Laboratory of Optical Astronomy, National Astronomical Observatories, Chinese Academy of Sciences, Beijing 100101, People's Republic of China}
\affiliation{School of Astronomy and Space Science, University of the Chinese Academy of Sciences, Beijing 100049, People's Republic of China}
\email{chenxiaodian@nao.cas.cn}

\author[orcid=0009-0005-0485-418X]{Li-Heng Wang}
\affiliation{Shandong Key Laboratory of Space Environment and Exploration Technology, Institute of Space Sciences, School of Space Science and Technology, Shandong University, Weihai, Shandong 264209, People's Republic of China}
\email{243786559@qq.com}

\begin{abstract}

Eclipsing binaries are crucial astrophysical laboratories for studying stellar parameters and evolutionary processes. In this study, we constructed a machine-learning-based model for systematic phenomenological classification of eclipsing binaries. We first extracted eclipsing binaries from the ASAS-SN variable star catalog and cross-matched them with TESS targets. The corresponding TESS light curves were processed through a unified pipeline, resulting in a high-quality training set of 9576 eclipsing binary light curves (2801 EA, 1930 EB, and 4845 EW systems). We designed and trained a fully connected neural network (FCNN) that achieved accuracy of 99.23\% and 99.03\% on the validation and test set respectively, demonstrating excellent performance. Applying the trained neural network to a total of 20196 TESS eclipsing binaries collected from multiple star catalogs and performing manual visual inspection, we finally obtained 13376 EA, 2114 EB, and 4706 EW systems. The standardized preprocessing pipeline and high-performance classifier developed in this study provide a reliable tool for the rapid automated classification of massive numbers of eclipsing binary in future photometric surveys.

\end{abstract}


\section{Introduction} 

Eclipsing binaries are binary systems in which two stars orbit each other and experience at least one observable eclipse. Based on the appearance of their light curves, eclipsing binaries are typically divided into three classes: EA, EB, and EW. EA-type binaries are marked by approximately constant out-of-eclipse brightness and distinct, abrupt eclipses that occupy only a small fraction of the light curve. EB-type binaries exhibit continuously variable light curves and a large difference in the depths of their two minima. EW-type binaries also have continuously variable light curves, but their two minima are usually similar in depth \citep{Kallrath2009Eclipsing}. Analyzing eclipsing binary light curves allows for the determination of relative stellar parameters and key orbital elements such as relative radii, temperature ratio, and orbital inclination. However, absolute fundamental parameters including stellar masses and radii require radial velocity measurements from double-lined spectroscopic binary (SB2) systems \citep{1991A&ARv...3...91A}.  The study of such systems offers profound insights into single-star evolution and helps explain a wide range of observed astrophysical phenomena \citep{2021A&C....3600488C}.

In recent years, large-scale surveys such as 
the Optical Gravitational Lensing Experiment (OGLE; \citealt{2015AcA....65....1U}; \citealt{2016AcA....66..405S}; \citealt{2016AcA....66..421P}; \citealt{2024AcA....74..241G}), Kepler space telescope \citep{2010Sci...327..977B}, the All-Sky Automated Survey for Supernovae (ASAS-SN; \citealt{2018MNRAS.477.3145J}), Gaia \citep{2016A&A...595A...1G}, the Transiting Exoplanet Survey Satellite (TESS; \citealt{2015JATIS...1a4003R}), and the Zwicky Transient Facility (ZTF) survey (\citealt{2019PASP..131a8002B}; \citealt{2019PASP..131a8003M}; \citealt{2020ApJS..249...18C}) have greatly advanced eclipsing binary research. These surveys provide massive, long-term, and high-precision photometric data, forming a foundational resource for modern studies of eclipsing binaries. To fully leverage these resources, a crucial step is to phenomenologically classify eclipsing binaries into EA, EB, and EW types based on their light curve morphology. However, the sheer volume of data renders traditional manual inspection and classification methods based on simple rules inefficient and impractical. As a result, machine-learning-based automated eclipsing binary classification has become a major research direction in modern astrophysics.

Numerous studies have employed machine-learning methods for detection or classification of eclipsing binaries. \citet{2016MNRAS.456.2260A} applied self-organizing maps and random forests to classify variable stars and eclipsing binaries in K2 \citep{2014PASP..126..398H} fields 0-4.  \citet{2023MNRAS.520..828D} applied supervised learning—particularly a compound decision tree—to classify eclipsing binaries in the Vista Variables in the Via Lactea (VVV; \citealt{2010NewA...15..433M}) into detached, semi-detached, and contact classes. TESS enables large-scale, high-precision eclipsing binary discovery via its all-sky coverage, short-cadence photometry, and open data policy. \citet{2022ApJS..258...16P} proposed a morphology parameter to characterize eclipsing binaries from TESS data, based on a dimensionality reduction algorithm. \citet{2025PASP..137d4503S} identified 9351 new eclipsing binaries using 2-minute cadence TESS light curves based on a hybrid deep learning model, with the study focusing solely on the detection of eclipsing binaries without further classification. \citet{2025ApJS..279...50K} compiled a catalog of 10001 uniformly vetted and validated eclipsing binaries by screening TESS full-frame image data with a neural network, with its scope also limited to the discovery of eclipsing binaries rather than further classification. \citet{2025ApJS..276...57G} employed a random forest method to classify periodic variable stars observed by TESS into 12 subtypes, among which the eclipsing binaries were further divided into EA, EB, and EW types. \citet{2025ApJ...986...19W} used TESS data to construct a variable star classifier based on the random forest algorithm, but its further subdivision of eclipsing binaries was limited to EA and EW types.

Some of these studies of TESS eclipsing binaries are limited to identification, whereas others have progressed to classification. However, a key limitation of existing research lies in the imprecise phenomenological classification of EB subtypes. Therefore, developing a more robust, efficient, and reproducible automatic method for the classification of eclipsing binaries into EA, EB, and EW types remains of great scientific and practical value. This study aims to integrate existing eclipsing binary data resources derived from TESS and construct a fully connected neural network (FCNN) model (a classic multi-layer perceptron architecture popularized by the back-propagation algorithm; \citealt{1986Natur.323..533R}) to achieve automatic classification of eclipsing binaries. 

\section{Preparation of the FCNN Training Dataset} 

\subsection{Data sources}
\label{2.1.Data sources}
\begin{figure*}
\centering
\includegraphics[width=0.7\textwidth]{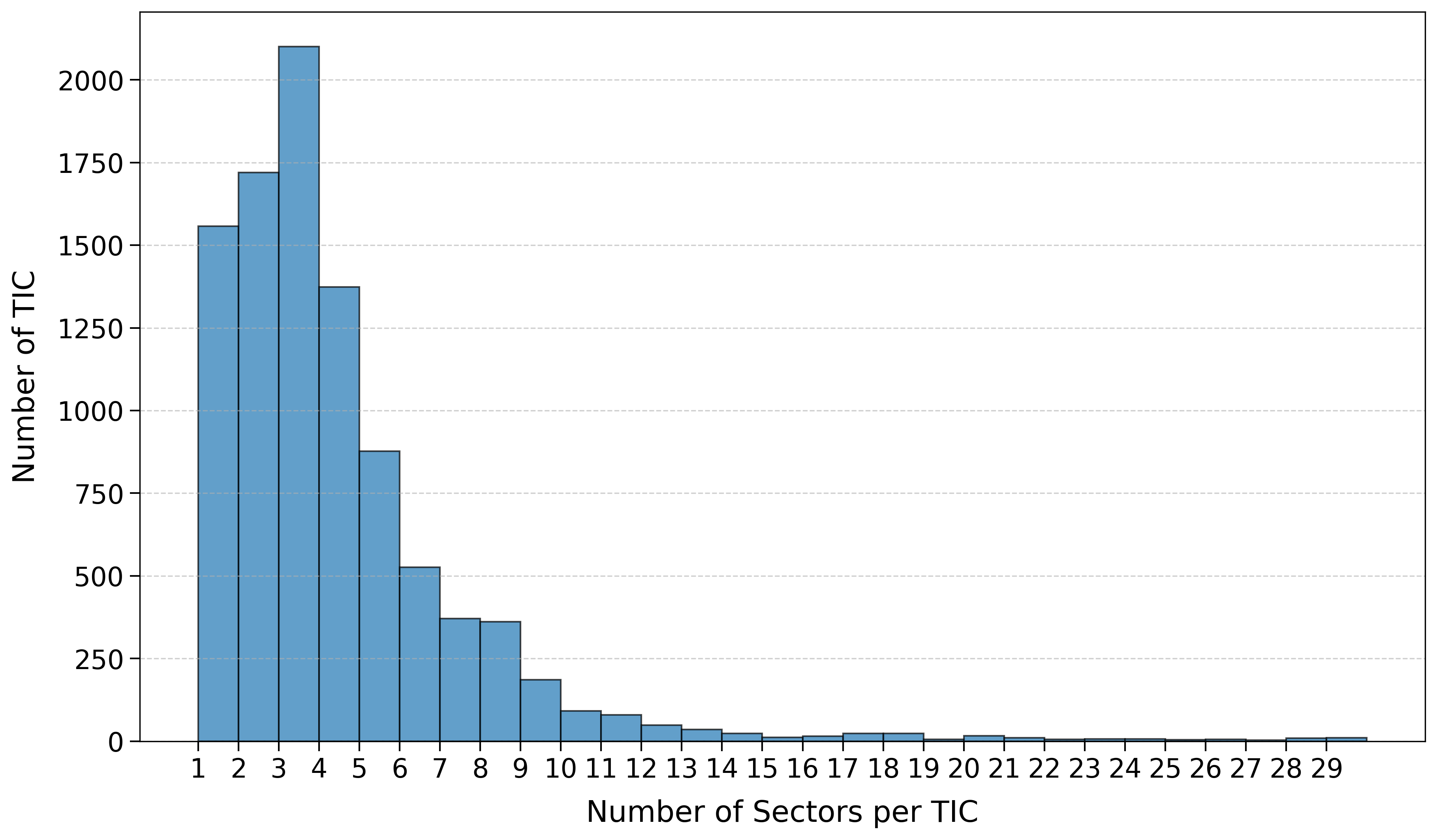}
\caption{A histogram of the number of available sectors per target.}
\label{fig:sector_hist}
\end{figure*}

We first extracted the eclipsing binaries from the ASAS-SN catalog, which itself provides a machine-learning-based classification into EA, EB, and EW types along with the corresponding class probability for each target. To ensure the quality of the data and the reliability of the classification, samples were filtered based on two criteria: the machine learning class probability should be no less than 0.99 and the Lafler-Kinman string length (\citealt{1965ApJS...11..216L}; \citealt{2002A&A...386..763C}) should be less than 0.1 \citep{2025ApJS..277...51L}. The filtered objects were cross-matched with TESS Input Catalog, identifying 3107 EA, 2633 EB, and 5147 EW systems. The TESS light curves for these objects were downloaded using the lightkurve package \citep{2018ascl.soft12013L} in Python. The average number of sectors per target is 4.44 for EA, 4.87 for EB, and 4.81 for EW. Figure \ref{fig:sector_hist} shows a histogram of the number of available sectors per target.

We employed high-level preprocessed light curves from the TESS mission and scientific teams, without performing aperture photometry directly from target pixel files (TPFs) or full-frame images (FFIs). For each target, we ranked all available TESS light curves using a hierarchical selection strategy. First, light curves were prioritized by exposure time in the order 1800\,s, 600\,s, 200\,s, 120\,s, as longer exposures generally yield a higher signal-to-noise ratio. For a small number of contact binaries with extremely short orbital periods (e.g., near $0.22\ \text{d}$), the 1800\,s cadence may cause phase smearing and artificially smooth eclipse features. However, such systems account for only a small fraction of the sample and do not significantly affect the overall classification results. For light curves with identical exposure times, we further ranked them by data processing pipeline in the priority order: SPOC, TESS-SPOC, TASOC, QLP, TGLC. The highest-ranked light curve was chosen as the optimal one for each target. If multiple sectors provided equally optimal data under the above criteria, we randomly selected one representative sector to avoid redundancy. For flux selection, we adopted the flux column following the priority: PDCSAP, SAP, and raw flux. The highest-priority available flux was used for the subsequent analysis.

\subsection{Light curve preprocessing}
\label{sec:Light curve preprocessing}
\begin{figure*}
\centering
\includegraphics[width=1\textwidth]{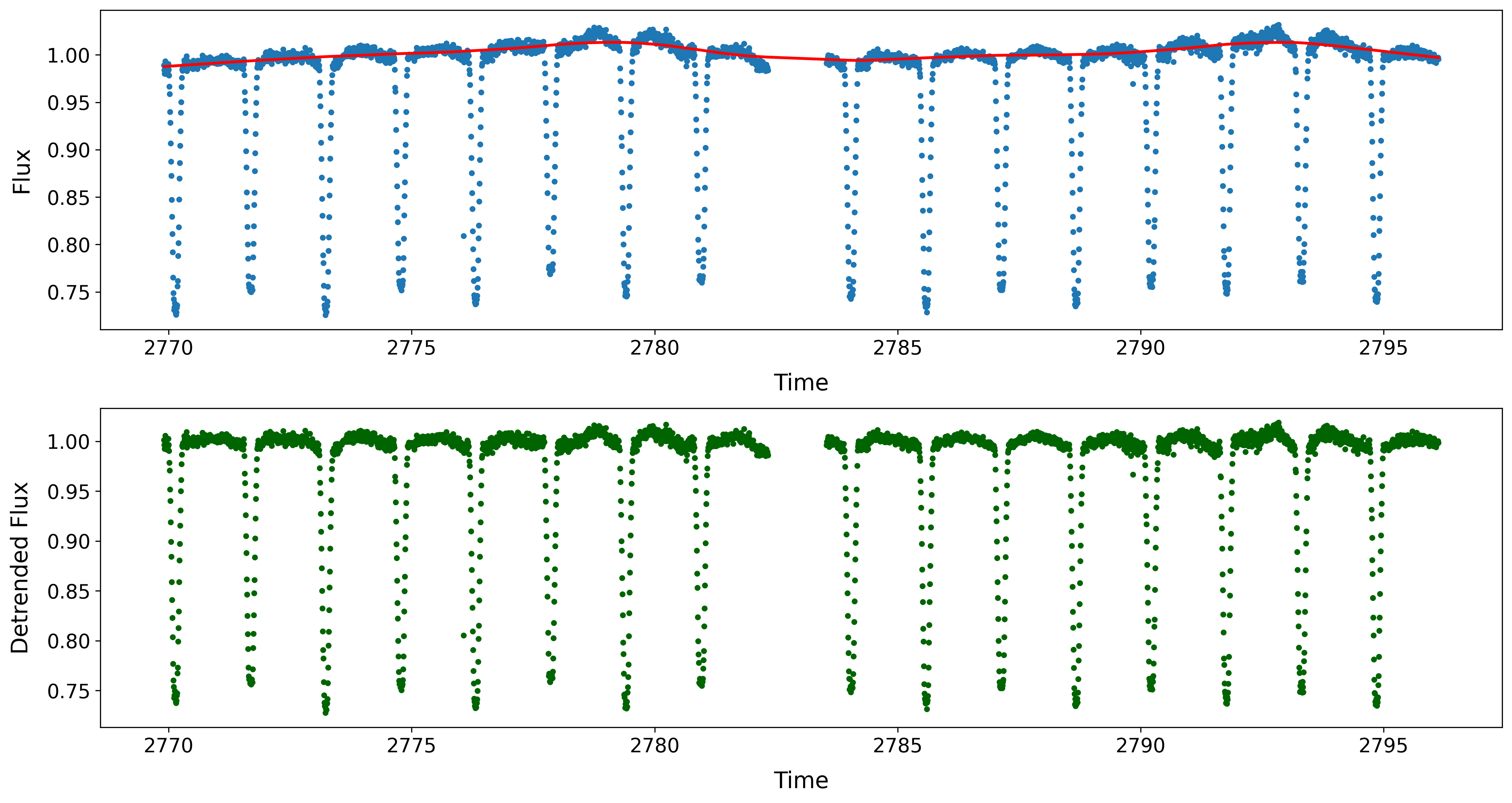}
\caption{Detrending process of the light curve for TIC 133966005. The upper panel shows the original light curve (blue dots) and the corresponding LOWESS-fitted trend (red line), while the lower panel displays the detrended light curve.}
\label{fig:detrending}
\end{figure*}

\begin{figure*}
\centering
\includegraphics[width=1\textwidth]{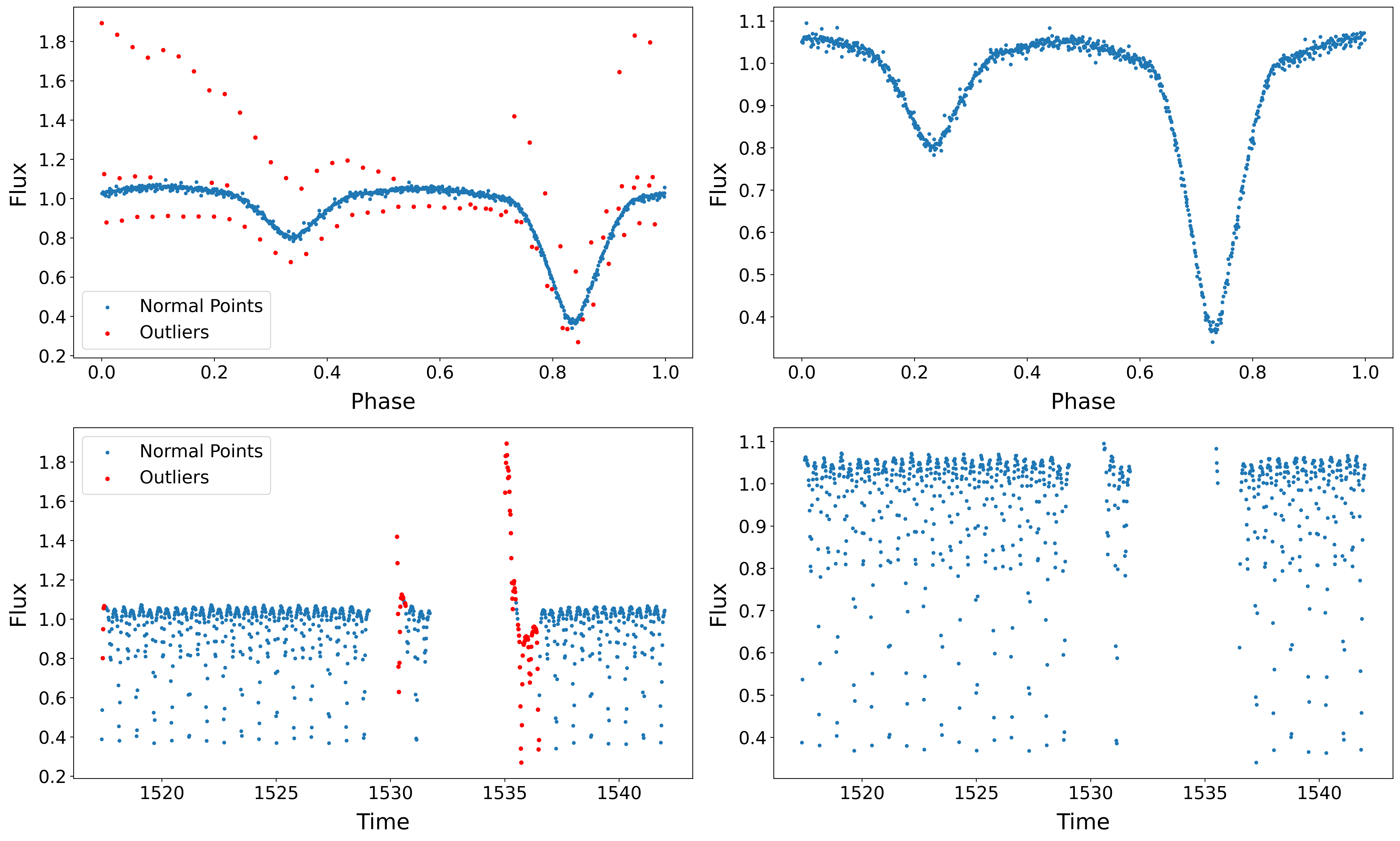}
\caption{Data denoising process of the light curve for TIC 391623240. The upper two panels display the denoising process in the phase domain, and the lower two panels show the corresponding results in the time domain. Blue dots represent valid retained data, while red dots represent the noise points removed during the denoising process.}
\label{fig:Data denoising}
\end{figure*}

All TESS light curves underwent unified preprocessing to ensure structural consistency of input data of the FCNN model:
\begin{enumerate}
\item Detrending: Locally weighted scatterplot smoothing (LOWESS;  
\citealt{Cleveland01121979}) was applied for light curve detrending. The smoothing parameter was adaptively determined based on stellar variability periods. The long-term systematic trend was fitted and removed to preserve intrinsic short-term flux variations. The upper panel of Figure \ref{fig:detrending} shows the original light curve (blue dots) and the corresponding LOWESS-fitted trend (red line) for TIC 133966005, while the lower panel displays the detrended light curve.
\item Phase folding: Light curves were phase-folded using periods provided in the ASAS-SN catalog, mapping time series to the phase interval $[0, 1]$.
\item Data denoising: First, global smoothing was performed on the phase-folded light curve using periodically extended LOWESS. The phase of maximum flux on the smoothed curve was defined as the new zero phase, and the light curve was phase-shifted to roughly place the two eclipses near phases 0.25 and 0.75. This preliminary adjustment cannot ensure the eclipses are precisely located at these two phases, particularly for systems with flat out-of-eclipse baselines or eccentric orbits. Local Gaussian fitting was then performed around the initial eclipse estimates, where each Gaussian center was set as a free parameter instead of being fixed at 0.25 and 0.75 to identify the eclipse regions.  Different LOWESS smoothing strengths were used to fit the eclipse and non-eclipse regions separately. Based on the fitted curves, up to 10 iterations of IQR (Interquartile Range) filtering were adopted to gradually remove noise points. As shown in Figure \ref{fig:Data denoising}, the upper two panels display the denoising process of TIC 391623240 in the phase domain, and the lower two panels show the corresponding results in the time domain. Blue dots represent valid retained data, while red dots represent the noise points removed during the denoising process.
\item Normalization: Each light curve was linearly normalized to the interval $[0, 1]$. The normalized flux $f_i'$ was calculated from the original flux values as: 
\begin{equation}
f_i'=\frac{f_i-f_{\mathrm{min}}}{f_{\mathrm{max}}-f_{\mathrm{min}}},\quad i=1,2,\dots,n
\end{equation}
where $f_{\mathrm{min}}$ and $f_{\mathrm{max}}$ 
are the minimum and maximum flux values respectively.
\item Primary minimum alignment: A double Gaussian model was adopted to fit each light curve. By comparing the fitted flux values at the two means of the Gaussian functions, the minimum with the smaller flux was identified as the primary minimum. Finally, the entire light curve was phase-shifted to align the primary minimum at phase 0. The phase range was adjusted to $[-0.75, 0.25]$ to avoid splitting one of the eclipses across the phase boundary, which could negatively impact the neural network training process. This range ensures that, for most combinations of orbital parameters, both eclipses remain far from the boundaries.
\item Binning procedure: Aligned light curves were divided into 500 equal-width bins, and the median flux of each bin was used as its representative value. The resulting light curve feature vectors retained the periodic shape characteristics of the original light curves while meeting the fixed input dimension requirement of the neural network, enabling training of all samples in a unified feature space.
\end{enumerate}

\subsection{Training set construction}
\label{sec:Training set construction}
\begin{figure*}
\centering
\includegraphics[width=1\textwidth]{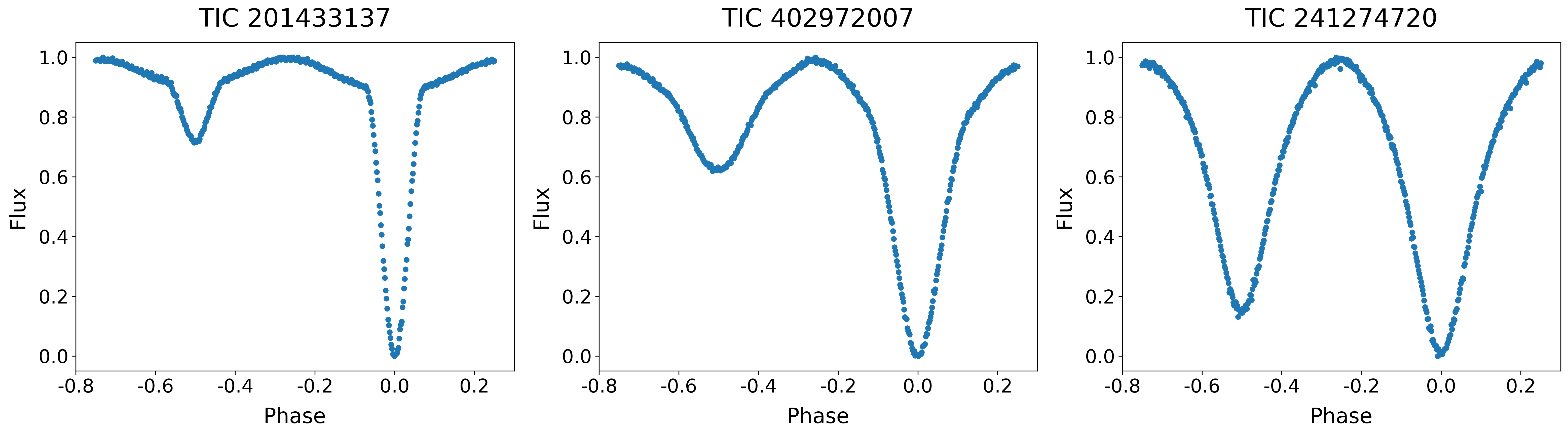}
\caption{Representative light curves from left to right for EA, EB, and EW systems.}
\label{fig:3_lc}
\end{figure*}

After all preprocessing steps, manual visual inspection was performed on these light curves to reclassify clearly misclassified samples into their correct classes, thereby ensuring the quality of the training set and increasing the sample size. The final neural network training set comprises 2801 EA, 1930 EB, and 4845 EW systems. Representative light curves from left to right for EA, EB, and EW systems are shown in Figure \ref{fig:3_lc}.

\section{Neural Network Model}
\subsection{Model architecture}
This study employed a FCNN model with a simple structure for the classification of light curves. To mitigate class imbalance among EA, EB, and EW systems, we employed a weighted cross-entropy loss function \citep{Goodfellow2016Deep}, with class weights determined based on the class distribution in the dataset. The input layer dimension is 500, corresponding to the number of data points in a single preprocessed light curve. The model consists of two hidden layers, composed of 128 and 64 neurons respectively, both using the ReLU \citep{10.5555/3104322.3104425} activation function. A Dropout \citep{10.5555/2627435.2670313} layer with a dropout rate of 0.3 is added after each hidden layer to suppress overfitting. The output layer has a dimension of 3 (corresponding to the three eclipsing binary types) and outputs the predicted probability for each class. The choice of this network architecture was guided by a trade-off between model complexity and generalization performance, aiming to achieve sufficient representational capacity while avoiding overfitting given the dataset size. We also tested several alternative configurations, including varying the number of hidden layers, neurons per layer, and dropout rates; however, these modifications did not yield significant improvements over the adopted architecture and in some cases even led to worse performance.

\subsection{Training configuration}
\begin{figure*}
\centering
\includegraphics[width=1\textwidth]{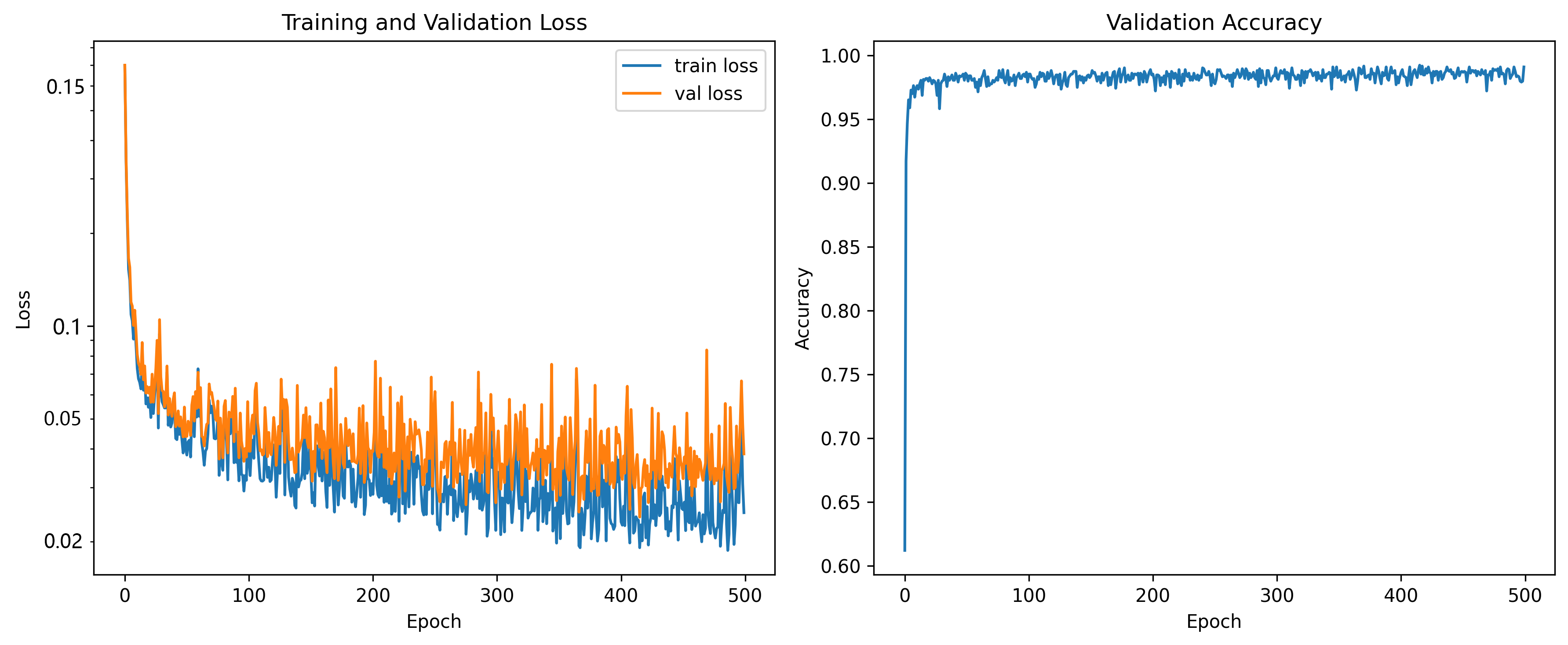}
\caption{Training dynamics of the FCNN model over 500 epochs. Left: Training and validation loss (logarithmic y-axis). Right: Validation accuracy. The x-axis denotes the epoch number.}
\label{fig:FNN_loss}
\end{figure*}

The light curves were subjected to stratified sampling with a 70\% : 15\% : 15\% split for the training, validation, and test sets, respectively. The model was trained using the Adam \citep{2014arXiv1412.6980K} optimizer with a learning rate of 0.0001. The model was trained for 500 epochs, and the model achieving the highest validation accuracy was saved. The optimal model attained a classification accuracy of 99.23\% on the validation set, indicating that it can effectively distinguish among the EA, EB, and EW types. Figure \ref{fig:FNN_loss} shows the training loss, validation loss (left panel), and validation accuracy (right panel) over 500 epochs.

\subsection{Model performance}
\begin{figure*}
\centering
\includegraphics[width=0.4\textwidth]{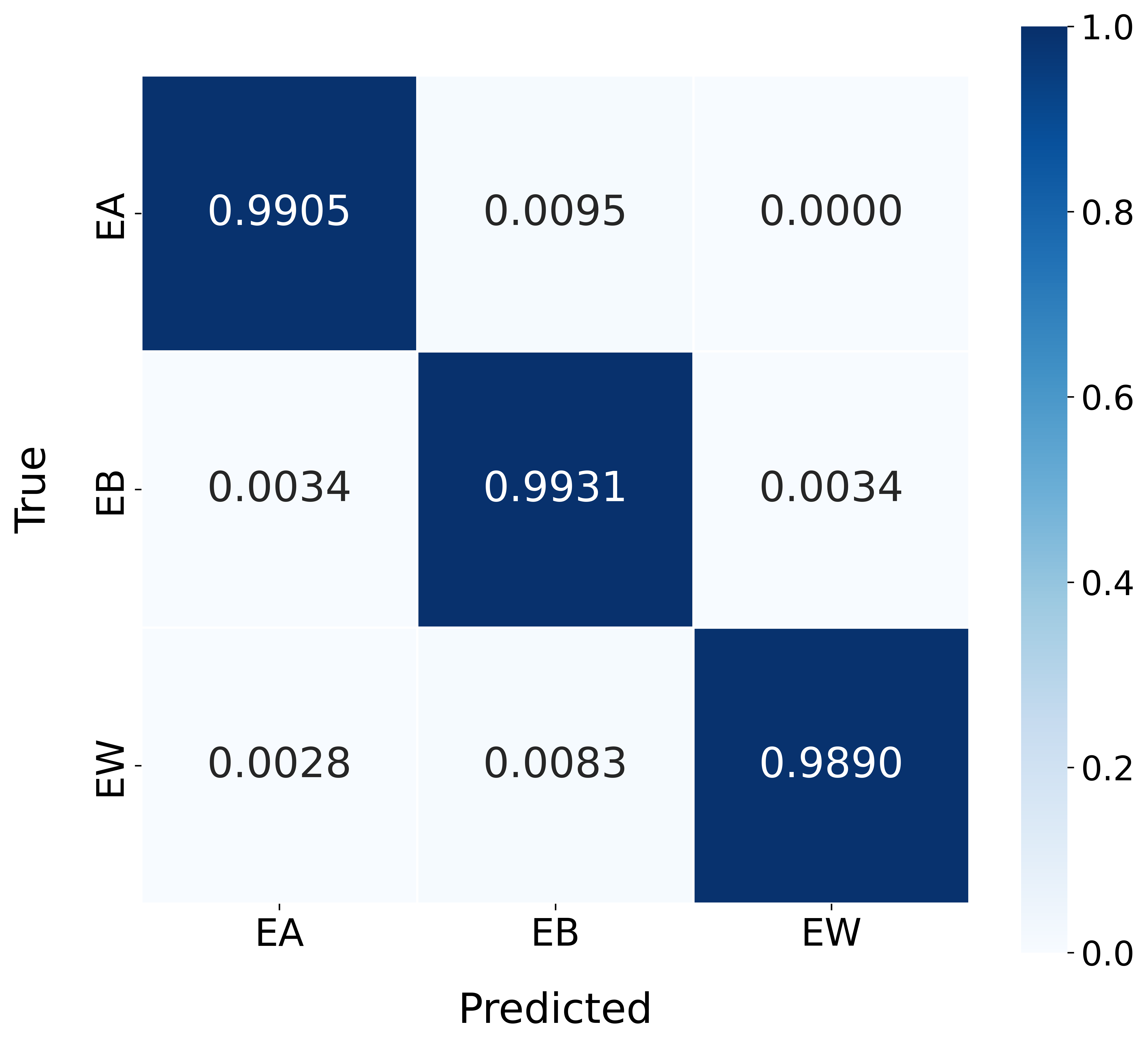}
\caption{Confusion matrix of the FCNN model evaluated on the test set.
\label{fig:FNN_cm}}
\end{figure*}

After training, the model was evaluated on the test set, achieving an accuracy of 99.03\%. Figure \ref{fig:FNN_cm} shows the confusion matrix of the model on the test set. Overall, the model demonstrates strong capability to distinguish between the three types of eclipsing binaries. The model's simple architecture and fast training/prediction capabilities make it highly efficient for processing large-scale eclipsing binary light curve data.

\section{Application} 
This section presents the application of the trained FCNN model to a large-scale TESS eclipsing binary dataset, aiming to automatically classify their light curves into EA, EB, and EW subtypes. The workflow comprises three main steps: (1) downloading and preprocessing light curves of the application set, (2) performing anomaly detection via an autoencoder \citep{2006Sci...313..504H} to filter out low-quality or non-eclipsing binary samples, and (3) performing classification using the FCNN model. The reliability of the classification results was then validated through manual inspection and cross-matching with existing catalogs.
\subsection{Application samples}
We integrated multiple published TESS eclipsing binary catalogs, including \citet{2022ApJS..258...16P}, \citet{2024AJ....167..192D}, \citet{2025AJ....169..202D}, \citet{2025ApJS..276...57G}, \citet{2025PASP..137d4503S}, \citet{2025ApJ...986...19W}, and \citet{2025ApJS..279...50K}, obtaining a total of 27777 eclipsing binaries. This serves as the application set for classifying eclipsing binaries into EA, EB, and EW subtypes. For each target, we downloaded a single light curve following the same selection criteria as described in Section \ref{2.1.Data sources}, ensuring full consistency in data preparation. For targets lacking a period in  \citet{2025AJ....169..202D} and \citet{2025PASP..137d4503S}'s catalogs, the Box Least Squares (BLS; \citealt{2002A&A...391..369K}) method was employed for period search. An initial scan over a coarse period grid was performed to obtain an approximate period, followed by a refined search using a denser grid around this estimate to accurately determine the optimal period.

To serve as new application data for the FCNN model, all downloaded light curves underwent the same preprocessing workflow as described in Section \ref{sec:Light curve preprocessing}, including detrending, phase folding, data denoising, normalization, primary minimum alignment and binning procedure. After preprocessing, we observed a large number of light curves with poor data quality or obvious non-eclipsing binary characteristics. To address this issue, an autoencoder was further trained to detect abnormal samples that deviated significantly from the shape features of eclipsing binary light curves, thereby filtering out low-quality or non-eclipsing binary samples.

\subsection{Anomaly detection with Autoencoder}
The core idea of the autoencoder is to unsupervisedly learn the intrinsic features of normal samples, enabling the model to effectively reconstruct normal data while producing significantly larger reconstruction errors for abnormal samples. Thus, the reconstruction error serves as a metric to quantify the severity of the anomaly. The autoencoder was trained on the labeled high-quality light curves from multiple TESS sectors as described in Section \ref{sec:Training set construction}, including 2801 EA, 1930 EB, and 4845 EW samples. They were split into training and validation sets by stratified sampling, with 80\% for training and 20\% for validation. The input dimension was 500 (corresponding to the 500 data points of the light curve), and the dimension of  the bottleneck layer was set to 32. The model was trained by minimizing the mean squared error (MSE) between the input and the reconstructed output, with an early stopping strategy to prevent overfitting.

\begin{figure*}
\plotone{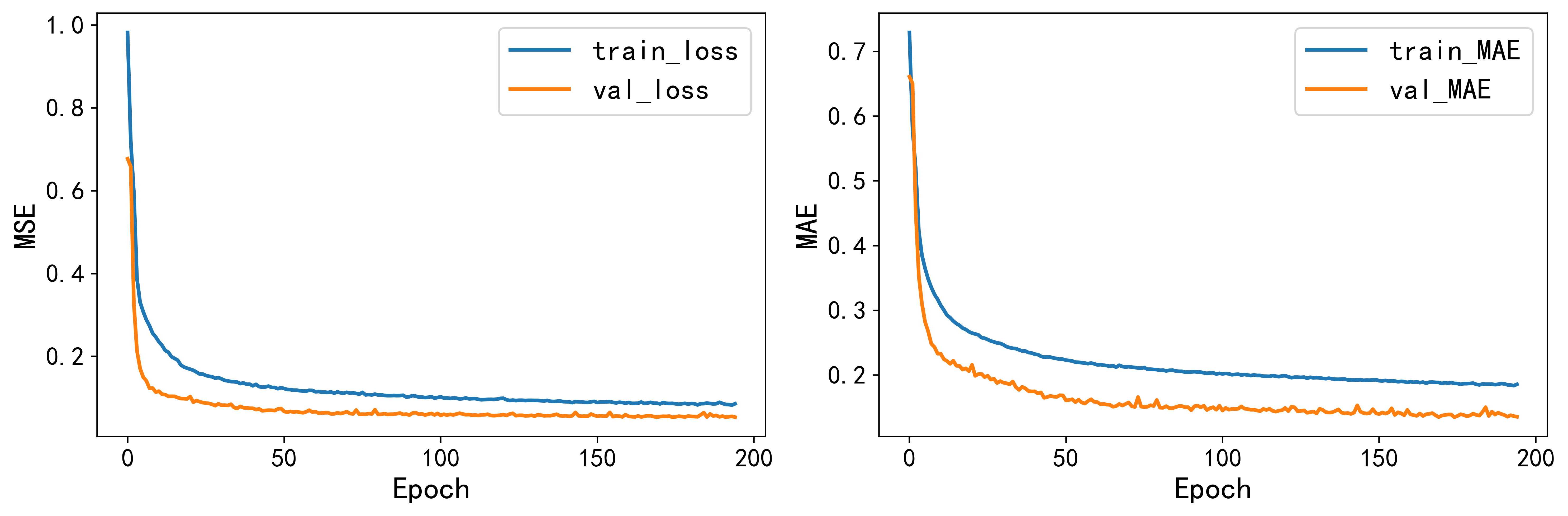}
\caption{Autoencoder training curves: loss and MAE versus training epochs. Left panel presents the training and validation loss curves, while the right panel shows the training and validation MAE curves. 
\label{fig:autoencoder_train}}
\end{figure*}

Figure \ref{fig:autoencoder_train} shows the autoencoder training dynamics, plotting the loss and mean absolute error (MAE) as a function of the number of training epochs. The left panel shows the loss curves, with the blue and orange curves corresponding to the training and validation loss respectively. The right panel shows the MAE curves, with the blue and orange curves corresponding to the training and validation MAE respectively. This phenomenon, in which the validation loss and MAE are lower than those of the training set, can be attributed to Dropout regularization. During training, Dropout randomly deactivates neurons, increasing task difficulty and training loss. During validation, it is disabled, allowing full network utilization and resulting in lower loss and MAE. All four curves decreased rapidly in the early training stage and gradually stabilized, indicating that the autoencoder effectively learned the main structural features of the light curves and successfully reconstructed them. This provides a stable and reliable foundation for subsequent anomaly detection based on reconstruction errors.

After training the autoencoder, reconstruction errors were calculated for all training light curves and new application light curves. A One-Class Support Vector Machine (One-Class SVM; \citealt{10.1162/089976601750264965}) model was trained on these reconstruction errors to determine a robust anomaly detection boundary of the autoencoder. Specifically, a One-Class SVM model with a radial basis function (RBF; \citealt{broomhead1988multivariable}) kernel was trained using the reconstruction errors of the training set as normal samples. The core goal of this model is to learn the support boundary of normal reconstruction errors in the high-dimensional space. For new samples, their reconstruction errors were input into the trained One-Class SVM model. Samples whose errors fell within the learned support boundary were classified as normal, while those outside were considered abnormal. The result showed that 71.66\% new samples were classified as normal by the One-Class SVM model. Therefore, the 70th percentile of reconstruction errors was adopted as the final threshold for autoencoder-based anomaly detection. Light curves below this threshold were retained as normal for subsequent classification. By combining the strength of the autoencoder in learning latent features with the ability of the SVM in determining the anomaly boundary, this method avoids the subjectivity of manual threshold selection and filters out abnormally shaped or heavily noise-contaminated light curves more accurately.

However, both normal and abnormal samples contained light curves with incorrect periods (e.g., the true period was twice or half the current period). To address this issue, the \texttt{scipy.signal.find\_peaks} function in Python was used to identify minima in each light curve. If one minimum was detected, the period was adjusted to twice the current value. If four minima were detected, the period was adjusted to half the current value. Finally, manual visual inspection was performed to correct samples with obviously incorrect periods and remove those with extremely poor data quality. In cases where an originally anomalous light curve was corrected by period adjustment and subsequently classified as normal, it was manually reassigned to the normal set. A total of 20196 eclipsing binaries were processed to the final classification stage.

\subsection{Classification results}
The 20196 eclipsing binaries were input into the FCNN model, which output the predicted probability of each light curve belonging to EA, EB, or EW. For targets with two light curves, the class with a higher predicted probability was selected as the final classification result. The application results of the FCNN were 13241 EA, 2165 EB, and 4790 EW eclipsing binaries. Subsequent manual verification of all applications yielded the following breakdown, as summarized in Table \ref{tab:classification}.

\begin{table}[htbp]
\centering
\caption{Classification results after manual verification}  
\begin{tabular}{c|c|ccc|c}  
\hline
\multirow{2}{*}{Predicted Class} & \multirow{2}{*}{Total} & \multicolumn{3}{c|}{True Class} & \multirow{2}{*}{Precision} \\  
\cline{3-5}  
 & & EA & EB & EW & \\  
\hline
EA & 13241 & 13077 & 105 & 59 & 98.76\% \\  
\hline
EB & 2165 & 181 & 1957 & 27 & 90.39\% \\
\hline
EW & 4790 & 118 & 52 & 4620 & 96.45\% \\
\hline
\end{tabular}
\label{tab:classification}  
\end{table}

After manual correction, the final classification results consisted of 13376 EA, 2114 EB, and 4706 EW, totaling 20196 eclipsing binaries. We cross-matched our final eclipsing binary catalog with the Gaia DR3 catalog \citep{2016A&A...595A...1G} using a matching radius of $3\ \text{arcsec}$. A representative excerpt of our catalog is presented in Table \ref{tab:catalog}, showing the first ten eclipsing binaries with their key parameters including ID, celestial coordinates (RA and DEC), eclipsing binary type (Type), orbital period (Period), the epoch of the primary minimum ($T_0$) and its associated uncertainty ($T_0$ err), effective temperature (Teff), surface gravity (logg), TESS magnitude (Tmag), BP-RP color (BP-RP), BP-RP color excess (E(BP-RP)), dereddened BP-RP color ($(BP-RP)_0$), parallax (Plx), G-band magnitude (Gmag), extinction in G-band (AG), the absolute G-band magnitude ($M_G$), observing sector (Sector), data processing pipeline (Pipeline), cadence, and flux type (Flux).

\begin{deluxetable*}{ccccc>{\bfseries}c>{\bfseries}ccccccccccccccc}
\tablewidth{0pt}
\setlength{\tabcolsep}{0.8pt}
\tablecaption{TESS eclipsing binary catalog \label{tab:catalog}}
\tablehead{
\colhead{TIC} & \colhead{RA} & \colhead{DEC} & \colhead{Type} & \colhead{Period} & \colhead{$T_0$} & \colhead{$T_0$ err} & \colhead{Teff} & \colhead{logg} & \colhead{Tmag} & \colhead{BP-RP} & \colhead{E(BP-RP)} & \colhead{$(BP-RP)_0$} & \colhead{Plx} & \colhead{Gmag} & \colhead{AG} & \colhead{$M_G$} & \colhead{Sector} & \colhead{Pipeline} & \colhead{Cadence} & \colhead{Flux} \\
\colhead{} & \colhead{(deg)} & \colhead{(deg)} & \colhead{} & \colhead{(days)} & \colhead{(BJD-2457000)} & \colhead{(days)} & \colhead{(K)} & \colhead{} & \colhead{(mag)} & \colhead{(mag)} & \colhead{(mag)} & \colhead{(mag)} & \colhead{(mas)} & \colhead{(mag)} & \colhead{(mag)} & \colhead{(mag)} & \colhead{} & \colhead{} & \colhead{(s)} & \colhead{}
}
\startdata
8636 & 219.00555 & -27.56944 & EA & 3.886554 & 1603.714099 & 0.025718 & 4841 & 4.198 & 12.355 & 1.2199 & 0.1081 & 1.1118 & 2.6452 & 12.9693 & 0.2026 & 4.879 & 11 & TESS-SPOC & 1800 & pdcsap \\
10507 & 219.04302 & -24.33205 & EA & 1.939769 & 1600.702973 & 0.000472 & 7469 & 3.879 & 11.479 & 0.5266 & 0.1999 & 0.3267 & 1.1286 & 11.7437 & 0.3719 & 1.6345 & 11 & TESS-SPOC & 1800 & pdcsap \\
49779 & 220.55332 & -24.2719 & EA & 9.392225 & 2340.061173 & 0.015707 &  &  & 14.567 & 2.9021 & 0 & 2.9021 & 12.1458 & 15.8807 & 0 & 11.3028 & 38 & TESS-SPOC & 600 & pdcsap \\
52357 & 220.4882 & -28.06576 & EB & 0.548154 & 1600.582514 & 0.000454 &  &  & 5.753 & 1.5612 & 0 & 1.5612 & 4.2282 & 6.5346 & 0 & -0.3346 & 11 & TESS-SPOC & 1800 & pdcsap \\
92193 & 221.88302 & -24.20143 & EW & 0.460555 & 1597.180943 & 0.001922 & 15001 & 3.432 & 10.764 & -0.0403 & 0.1794 & -0.2197 & 0.5315 & 10.6818 & 0.3336 & -1.0243 & 11 & QLP & 1800 & sap \\
101462 & 222.05765 & -24.61321 & EB & 8.347963 & 1606.490664 & 0.02002 &  &  & 9.831 & 0.8746 & 0 & 0.8746 & 9.443 & 10.2991 & 0 & 5.1746 & 11 & TESS-SPOC & 1800 & pdcsap \\
120016 & 222.57733 & -24.42774 & EW & 0.326371 & 1600.358677 & 0.000527 &  &  & 7.512 & 0.8957 & 0 & 0.8957 & 16.963 & 8.0829 & 0 & 4.2304 & 11 & TESS-SPOC & 1800 & pdcsap \\
627436 & 73.04731 & -25.19408 & EW & 0.578906 & 1438.296628 & 0.000088 & 7350 & 4.018 & 11.716 & 0.4301 & 0.041 & 0.3891 & 1.0842 & 11.9166 & 0.0767 & 2.0154 & 5 & TESS-SPOC & 1800 & pdcsap \\
665581 & 73.2857 & -25.84417 & EW & 0.471165 & 1438.458012 & 0.000215 & 5114 & 4.377 & 12.269 & 1.0416 & 0.0749 & 0.9667 & 3.232 & 12.827 & 0.1392 & 5.2351 & 5 & TESS-SPOC & 1800 & pdcsap \\
671564 & 73.40884 & -29.11019 & EW & 0.385623 & 1438.510568 & 0.000079 & 5749 & 4.239 & 11.665 & 0.8652 & 0.0015 & 0.8637 & 2.3347 & 12.0872 & 0.0029 & 3.9255 & 5 & TESS-SPOC & 1800 & pdcsap \\
\enddata
\tablecomments{This table is available in its entirety in machine-readable form in the online article.}
\end{deluxetable*}

\begin{figure*}
\centering
\includegraphics[width=1.0\textwidth]{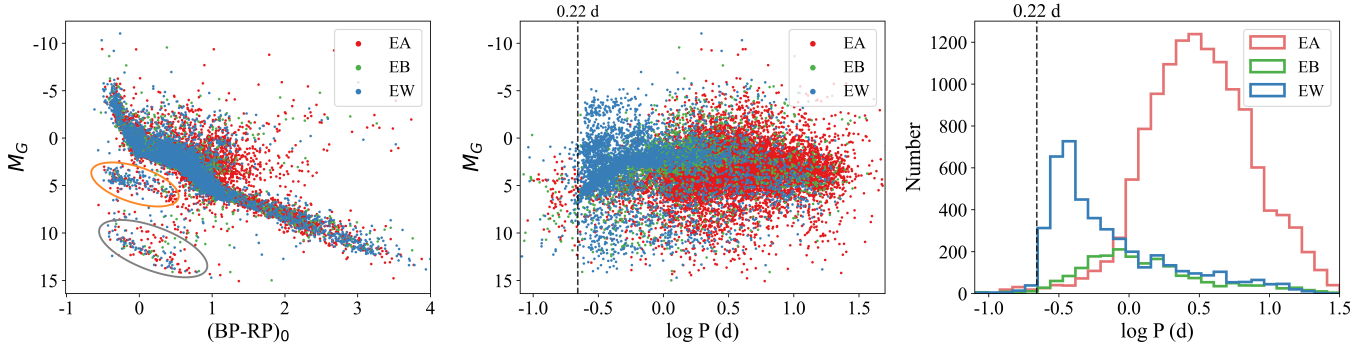}
\caption{Color–magnitude diagram, period–absolute-magnitude diagram, and period histogram for the eclipsing binary catalog. In the left panel, the orange elliptical area contains stars with extremely high temperatures but moderate luminosities, possibly corresponding to sdB or sdO subdwarfs in special evolutionary stages. The gray elliptical area is occupied by the white dwarf sequence, whose members remain very faint despite their high temperatures due to their small radii.}
\label{fig:three_diagram}
\end{figure*}

 Figure \ref{fig:three_diagram} presents the color-magnitude diagram, period–absolute-magnitude diagram, and period histogram for our eclipsing binary catalog. The left panel shows a color-magnitude diagram, where the vast majority of eclipsing binaries are located along the main sequence, extending from the upper-left high-temperature, high-luminosity region to the lower-right low-temperature, low-luminosity region. The orange elliptical area in the figure contains stars with extremely high temperatures but moderate luminosities, possibly corresponding to sdB or sdO subdwarfs in special evolutionary stages. The gray elliptical area in the figure is occupied by the white dwarf sequence, whose members remain very faint despite their high temperatures due to their small radii. The middle panel presents the period–absolute-magnitude diagram, illustrating the relationship between orbital period and luminosity, which shows a similar distribution of eclipsing binaries to that in \citet{2025ApJS..276...57G}. The right panel displays the period histogram. As seen in both the middle and right panels, EA-type systems span a wide range of periods, whereas EW-type systems are concentrated at shorter periods, with a sharp cutoff near $0.22\ \text{d}$ (marked by the vertical dashed line). This cutoff is consistent with the short-period limit of contact binaries (\citealt{1992AJ....103..960R}; \citealt{2012MNRAS.421.2769J}; \citealt{2015AJ....150..117Q}; \citealt{2019MNRAS.485.4588L}; \citealt{2020AJ....159..189L}).

\subsection{Comparison with other catalogs} 
We cross-matched our classification results with the TESS eclipsing binary catalog of \citet{2025ApJS..276...57G}, and the results are summarized in Table \ref{tab:cross_Gao}. The comparison indicates a high consistency for EA and EW classifications between our model and \citet{2025ApJS..276...57G}'s catalog. For EB, however, the higher confusion rate with EA and EW is attributed to the ambiguous inter-class boundaries inherent to the morphological characteristics of their light curves.

\begin{table}[htbp]
\centering
\caption{Cross-match results with \citet{2025ApJS..276...57G}'s catalog}
\begin{tabular}{c|c|ccc|c}
\hline
\multirow{2}{*}{Our Catalog} & \multicolumn{4}{c|}{\citet{2025ApJS..276...57G}'s Catalog} & \multirow{2}{*}{Rate} \\
\cline{2-5}
& Matched & EA & EB & EW & \\
\hline
EA (13376) & 3089 & 3074 & 13 & 2 & 99.51\% \\
\hline
EB (2114) & 564 & 202 & 285 & 77 & 50.53\% \\
\hline
EW (4706) & 1402 & 98 & 90 & 1214 & 86.59\% \\
\hline
\end{tabular}
\label{tab:cross_Gao}
\end{table}

\begin{figure*}
\centering
\includegraphics[width=0.5\textwidth]{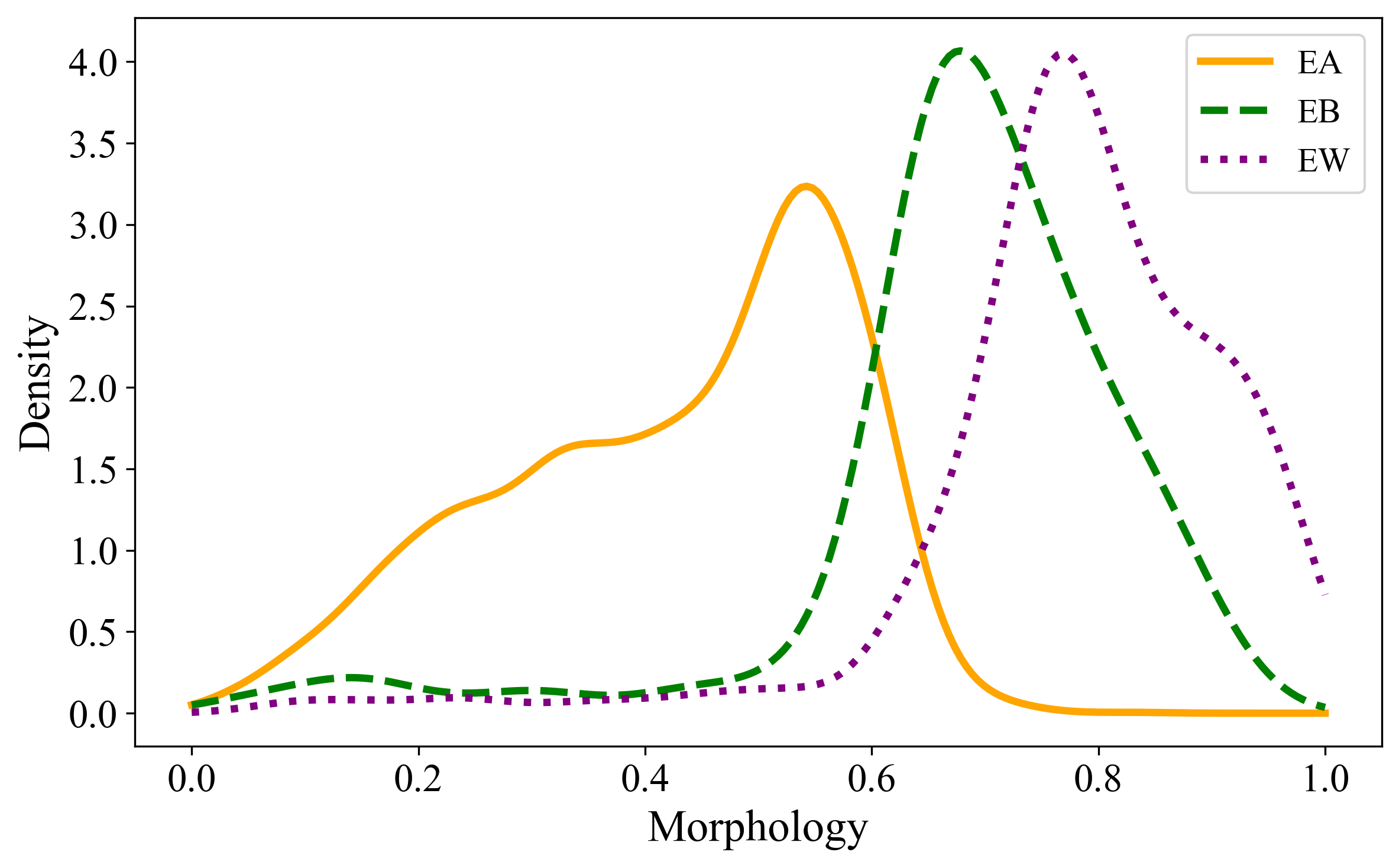}
\caption{Kernel density distribution curves of morphological parameters for EA, EB, and EW identified in this work, derived from cross-matching with the catalog by \citet{2022ApJS..258...16P}. The x-axis denotes the morphological parameter, and the y-axis represents the kernel density. 
\label{fig:Morphology}}
\end{figure*}

To further validate the reliability of our classification results, cross-matching was performed with the TESS eclipsing binary catalog provided by \citet{2022ApJS..258...16P}, and kernel density distribution curves of morphological parameters for matched objects were plotted in Figure \ref{fig:Morphology}. This figure clearly shows systematic differences in geometric separation among the three types. The primary peaks of the morphological parameter occur around 0.5 for EA, 0.7 for EB, and 0.8 for EW. This distribution trend aligns with the general geometric features of the three types: as the morphology parameter increases from 0 to 1, the geometric characteristics transition from detached, through semi-detached, to contact systems, which in turn corresponds to the variation of light curves from EA, through EB, to EW. The significant separation between the three curves demonstrates high consistency between our classification results and the morphological parameters from \citet{2022ApJS..258...16P}’s catalog.

\section{Conclusions}
This study developed an automated classification model for eclipsing binaries to achieve high-precision identification of EA, EB, and EW types. The process began with establishing a standardized light curve preprocessing workflow to unify the feature space of all light curves. Subsequently, a simple and efficient FCNN model was constructed, which achieved near-perfect performance with accuracies of 99.23\% on the validation set and 99.03\% on the test set for classification into the three types. Furthermore, an autoencoder integrated with a One-Class SVM was implemented to effectively filter out light curves in the application set exhibiting anomalous shapes or severe noise. Finally, the utilization of the trained FCNN model on the application set, followed by manual visual inspection, resulted in a final catalog of 13376 EA, 2114 EB, and 4706 EW systems. Future work will focus on incorporating key stellar physical properties (e.g., orbital period, temperature ratio) into our model and adapting it to classify eclipsing binaries detected by other large-scale surveys, such as the Large Synoptic Survey Telescope (LSST; \citealt{2019ApJ...873..111I}).

\begin{acknowledgments}

We sincerely thank the referee for constructive comments and valuable suggestions, which have significantly improved the quality of this manuscript.
This work is supported by the National Natural Science Foundation of China (NSFC; No. 12273018), the Taishan Scholars Young Expert Program of Shandong Province, the Qilu Young Researcher Project of Shandong University, the Young Data Scientist Project of the National Astronomical Data Center, and by the Cultivation Project for LAMOST Scientific Payoff and Research Achievement of CAMS-CAS. The calculations in this work were carried out at the Supercomputing Center of Shandong University, Weihai.

This paper makes use of data collected by the TESS mission, which are publicly available from the Mikulski Archive for Space Telescopes (MAST). Funding for the TESS mission is provided by NASA's Science Mission directorate. We acknowledge the use of public TESS data from pipelines at the TESS Science Office and at the TESS Science Processing Operations Center. 

This paper makes use of data from ASAS-SN. ASAS-SN is funded in part by the Gordon and Betty Moore Foundation through grants GBMF5490 and GBMF10501 to the Ohio State University, and also funded in part by the Alfred P. Sloan Foundation grant G-2021-14192. 

This work has made use of data from the European Space Agency (ESA) mission Gaia (\url{https://www.cosmos.esa.int/gaia}), processed by the Gaia Data Processing and Analysis Consortium (DPAC, \url{https://www.cosmos.esa.int/web/gaia/dpac/consortium}). Funding for the DPAC has been provided by national institutions, in particular the institutions participating
in the Gaia Multilateral Agreement.
\end{acknowledgments}

\bibliography{sample701}{}

@ARTICLE{2016MNRAS.456.2260A,
       author = {{Armstrong}, D.~J. and {Kirk}, J. and {Lam}, K.~W.~F. and {McCormac}, J. and {Osborn}, H.~P. and {Spake}, J. and {Walker}, S. and {Brown}, D.~J.~A. and {Kristiansen}, M.~H. and {Pollacco}, D. and {West}, R. and {Wheatley}, P.~J.},
        title = "{K2 variable catalogue - II. Machine learning classification of variable stars and eclipsing binaries in K2 fields 0-4}",
      journal = {\mnras},
         year = 2016,
        month = feb,
       volume = {456},
       number = {2},
        pages = {2260-2272},
          doi = {10.1093/mnras/stv2836},
archivePrefix = {arXiv},
       eprint = {1512.01246},
 primaryClass = {astro-ph.SR},
       adsurl = {https://ui.adsabs.harvard.edu/abs/2016MNRAS.456.2260A}
}

@ARTICLE{2023MNRAS.520..828D,
       author = {{Daza-Perilla}, I.~V. and {Gramajo}, L.~V. and {Lares}, M. and {Palma}, T. and {Ferreira Lopes}, C.~E. and {Minniti}, D. and {Clari{\'a}}, J.~J.},
        title = "{Automated classification of eclipsing binary systems in the VVV Survey}",
      journal = {\mnras},
         year = 2023,
        month = mar,
       volume = {520},
       number = {1},
        pages = {828-838},
          doi = {10.1093/mnras/stad141},
archivePrefix = {arXiv},
       eprint = {2302.01200},
 primaryClass = {astro-ph.SR},
       adsurl = {https://ui.adsabs.harvard.edu/abs/2023MNRAS.520..828D}
}

@ARTICLE{2022ApJS..258...16P,
       author = {{Pr{\v{s}}a}, Andrej and {Kochoska}, Angela and {Conroy}, Kyle E. and {Eisner}, Nora and {Hey}, Daniel R. and {IJspeert}, Luc and {Kruse}, Ethan and {Fleming}, Scott W. and {Johnston}, Cole and {Kristiansen}, Martti H. and {LaCourse}, Daryll and {Mortensen}, Danielle and {Pepper}, Joshua and {Stassun}, Keivan G. and {Torres}, Guillermo and {Abdul-Masih}, Michael and {Chakraborty}, Joheen and {Gagliano}, Robert and {Guo}, Zhao and {Hambleton}, Kelly and {Hong}, Kyeongsoo and {Jacobs}, Thomas and {Jones}, David and {Kostov}, Veselin and {Lee}, Jae Woo and {Omohundro}, Mark and {Orosz}, Jerome A. and {Page}, Emma J. and {Powell}, Brian P. and {Rappaport}, Saul and {Reed}, Phill and {Schnittman}, Jeremy and {Schwengeler}, Hans Martin and {Shporer}, Avi and {Terentev}, Ivan A. and {Vanderburg}, Andrew and {Welsh}, William F. and {Caldwell}, Douglas A. and {Doty}, John P. and {Jenkins}, Jon M. and {Latham}, David W. and {Ricker}, George R. and {Seager}, Sara and {Schlieder}, Joshua E. and {Shiao}, Bernie and {Vanderspek}, Roland and {Winn}, Joshua N.},
        title = "{TESS Eclipsing Binary Stars. I. Short-cadence Observations of 4584 Eclipsing Binaries in Sectors 1-26}",
      journal = {\apjs},
         year = 2022,
        month = jan,
       volume = {258},
       number = {1},
          eid = {16},
        pages = {16},
          doi = {10.3847/1538-4365/ac324a},
archivePrefix = {arXiv},
       eprint = {2110.13382},
 primaryClass = {astro-ph.SR},
       adsurl = {https://ui.adsabs.harvard.edu/abs/2022ApJS..258...16P}
}

@ARTICLE{2025ApJ...986...19W,
       author = {{Wang}, Li-Heng and {Li}, Kai and {Gao}, Xiang and {Guo}, Ya-Ni and {Sun}, Guo-You},
        title = "{Using Machine Learning Method for Variable Star Classification Using the TESS Sectors 1{\textendash}57 Data}",
      journal = {\apj},
         year = 2025,
        month = jun,
       volume = {986},
       number = {1},
          eid = {19},
        pages = {19},
          doi = {10.3847/1538-4357/add159},
archivePrefix = {arXiv},
       eprint = {2504.00347},
 primaryClass = {astro-ph.SR},
       adsurl = {https://ui.adsabs.harvard.edu/abs/2025ApJ...986...19W}
}

@ARTICLE{2024AJ....167..192D,
       author = {{Ding}, Xu and {Song}, ZhiMing and {Wang}, ChuanJun and {Ji}, KaiFan},
        title = "{Detection of Contact Binary Candidates Observed By TESS Using the Autoencoder Neural Network}",
      journal = {\aj},
         year = 2024,
        month = may,
       volume = {167},
       number = {5},
          eid = {192},
        pages = {192},
          doi = {10.3847/1538-3881/ad3048},
archivePrefix = {arXiv},
       eprint = {2404.06424},
 primaryClass = {astro-ph.SR},
       adsurl = {https://ui.adsabs.harvard.edu/abs/2024AJ....167..192D}
}

@ARTICLE{2025AJ....169..202D,
       author = {{Ding}, Xu and {Ji}, KaiFan and {Cheng}, QiYuan and {Song}, ZhiMing and {Wang}, JinLiang and {Tian}, XueFen and {Wang}, ChuanJun},
        title = "{Detection of Semidetached Eclipsing Binaries from TESS}",
      journal = {\aj},
         year = 2025,
        month = apr,
       volume = {169},
       number = {4},
          eid = {202},
        pages = {202},
          doi = {10.3847/1538-3881/adb846},
archivePrefix = {arXiv},
       eprint = {2504.14612},
 primaryClass = {astro-ph.SR},
       adsurl = {https://ui.adsabs.harvard.edu/abs/2025AJ....169..202D}
}

@ARTICLE{2025ApJS..276...57G,
       author = {{Gao}, Xinyi and {Chen}, Xiaodian and {Wang}, Shu and {Liu}, Jifeng},
        title = "{Classification of Periodic Variable Stars from TESS}",
      journal = {\apjs},
         year = 2025,
        month = feb,
       volume = {276},
       number = {2},
          eid = {57},
        pages = {57},
          doi = {10.3847/1538-4365/ad9dd6},
archivePrefix = {arXiv},
       eprint = {2412.06175},
 primaryClass = {astro-ph.SR},
       adsurl = {https://ui.adsabs.harvard.edu/abs/2025ApJS..276...57G}
}

@ARTICLE{2025PASP..137d4503S,
       author = {{Shan}, Ying and {Chen}, Jing and {Zhang}, Zichong and {Wang}, Liang and {Zou}, Zhiqiang and {Li}, Min},
        title = "{Identifying Eclipsing Binary Stars with TESS Data Based on a New Hybrid Deep Learning Model}",
      journal = {\pasp},
         year = 2025,
        month = apr,
       volume = {137},
       number = {4},
          eid = {044503},
        pages = {044503},
          doi = {10.1088/1538-3873/adc5a2},
archivePrefix = {arXiv},
       eprint = {2504.15875},
 primaryClass = {astro-ph.SR},
       adsurl = {https://ui.adsabs.harvard.edu/abs/2025PASP..137d4503S}
}

@ARTICLE{2025ApJS..279...50K,
       author = {{Kostov}, Veselin B. and {Powell}, Brian P. and {Fornear}, Aline U. and {Di Fraia}, Marco Z. and {Gagliano}, Robert and {Jacobs}, Thomas L. and {de Lambilly}, Julien S. and {Durantini Luca}, Hugo A. and {Majewski}, Steven R. and {Omohundro}, Mark and {Orosz}, Jerome and {Rappaport}, Saul A. and {Salik}, Ryan and {Short}, Donald and {Welsh}, William and {Alexandrov}, Svetoslav and {da Silva}, Cledison Marcos and {Dunning}, Erika and {G{\"u}hne}, Gerd and {Huten}, Marc and {Hyogo}, Michiharu and {Iannone}, Davide and {Lee}, Sam and {Magliano}, Christian and {Sharma}, Manya and {Tarr}, Allan and {Yablonsky}, John and {Acharya}, Sovan and {Adams}, Fred and {Barclay}, Thomas and {Montet}, Benjamin T. and {Mullally}, Susan and {Olmschenk}, Greg and {Pr{\v{s}}a}, Andrej and {Quintana}, Elisa and {Wilson}, Robert and {Balcioglu}, Hasret and {Kruse}, Ethan and {The Eclipsing Binary Patrol Collaboration}},
        title = "{The TESS Ten Thousand Catalog: 10,001 Uniformly Vetted and Validated Eclipsing Binary Stars Detected in Full-frame Image Data by Machine Learning and Analyzed by Citizen Scientists}",
      journal = {\apjs},
         year = 2025,
        month = aug,
       volume = {279},
       number = {2},
          eid = {50},
        pages = {50},
          doi = {10.3847/1538-4365/ade2d8},
archivePrefix = {arXiv},
       eprint = {2506.05631},
 primaryClass = {astro-ph.SR},
       adsurl = {https://ui.adsabs.harvard.edu/abs/2025ApJS..279...50K}
}

@ARTICLE{2019MNRAS.485.4588L,
       author = {{Li}, Kai and {Xia}, Qi-Qi and {Michel}, Raul and {Hu}, Shao-Ming and {Guo}, Di-Fu and {Gao}, Xing and {Chen}, Xu and {Gao}, Dong-Yang},
        title = "{Contact binaries at the short period cut-off - I. Statistics and the first photometric investigations of 10 totally eclipsing systems}",
      journal = {\mnras},
         year = 2019,
        month = jun,
       volume = {485},
       number = {4},
        pages = {4588-4600},
          doi = {10.1093/mnras/stz715},
archivePrefix = {arXiv},
       eprint = {1903.04765},
 primaryClass = {astro-ph.SR},
       adsurl = {https://ui.adsabs.harvard.edu/abs/2019MNRAS.485.4588L}
}

@ARTICLE{2015JATIS...1a4003R,
       author = {{Ricker}, George R. and {Winn}, Joshua N. and {Vanderspek}, Roland and {Latham}, David W. and {Bakos}, G{\'a}sp{\'a}r {\'A}. and {Bean}, Jacob L. and {Berta-Thompson}, Zachory K. and {Brown}, Timothy M. and {Buchhave}, Lars and {Butler}, Nathaniel R. and {Butler}, R. Paul and {Chaplin}, William J. and {Charbonneau}, David and {Christensen-Dalsgaard}, J{\o}rgen and {Clampin}, Mark and {Deming}, Drake and {Doty}, John and {De Lee}, Nathan and {Dressing}, Courtney and {Dunham}, Edward W. and {Endl}, Michael and {Fressin}, Francois and {Ge}, Jian and {Henning}, Thomas and {Holman}, Matthew J. and {Howard}, Andrew W. and {Ida}, Shigeru and {Jenkins}, Jon M. and {Jernigan}, Garrett and {Johnson}, John Asher and {Kaltenegger}, Lisa and {Kawai}, Nobuyuki and {Kjeldsen}, Hans and {Laughlin}, Gregory and {Levine}, Alan M. and {Lin}, Douglas and {Lissauer}, Jack J. and {MacQueen}, Phillip and {Marcy}, Geoffrey and {McCullough}, Peter R. and {Morton}, Timothy D. and {Narita}, Norio and {Paegert}, Martin and {Palle}, Enric and {Pepe}, Francesco and {Pepper}, Joshua and {Quirrenbach}, Andreas and {Rinehart}, Stephen A. and {Sasselov}, Dimitar and {Sato}, Bun'ei and {Seager}, Sara and {Sozzetti}, Alessandro and {Stassun}, Keivan G. and {Sullivan}, Peter and {Szentgyorgyi}, Andrew and {Torres}, Guillermo and {Udry}, Stephane and {Villasenor}, Joel},
        title = "{Transiting Exoplanet Survey Satellite (TESS)}",
      journal = {Journal of Astronomical Telescopes, Instruments, and Systems},
         year = 2015,
        month = jan,
       volume = {1},
          eid = {014003},
        pages = {014003},
          doi = {10.1117/1.JATIS.1.1.014003},
       adsurl = {https://ui.adsabs.harvard.edu/abs/2015JATIS...1a4003R}
}

@ARTICLE{2010Sci...327..977B,
       author = {{Borucki}, William J. and {Koch}, David and {Basri}, Gibor and {Batalha}, Natalie and {Brown}, Timothy and {Caldwell}, Douglas and {Caldwell}, John and {Christensen-Dalsgaard}, J{\o}rgen and {Cochran}, William D. and {DeVore}, Edna and {Dunham}, Edward W. and {Dupree}, Andrea K. and {Gautier}, Thomas N. and {Geary}, John C. and {Gilliland}, Ronald and {Gould}, Alan and {Howell}, Steve B. and {Jenkins}, Jon M. and {Kondo}, Yoji and {Latham}, David W. and {Marcy}, Geoffrey W. and {Meibom}, S{\o}ren and {Kjeldsen}, Hans and {Lissauer}, Jack J. and {Monet}, David G. and {Morrison}, David and {Sasselov}, Dimitar and {Tarter}, Jill and {Boss}, Alan and {Brownlee}, Don and {Owen}, Toby and {Buzasi}, Derek and {Charbonneau}, David and {Doyle}, Laurance and {Fortney}, Jonathan and {Ford}, Eric B. and {Holman}, Matthew J. and {Seager}, Sara and {Steffen}, Jason H. and {Welsh}, William F. and {Rowe}, Jason and {Anderson}, Howard and {Buchhave}, Lars and {Ciardi}, David and {Walkowicz}, Lucianne and {Sherry}, William and {Horch}, Elliott and {Isaacson}, Howard and {Everett}, Mark E. and {Fischer}, Debra and {Torres}, Guillermo and {Johnson}, John Asher and {Endl}, Michael and {MacQueen}, Phillip and {Bryson}, Stephen T. and {Dotson}, Jessie and {Haas}, Michael and {Kolodziejczak}, Jeffrey and {Van Cleve}, Jeffrey and {Chandrasekaran}, Hema and {Twicken}, Joseph D. and {Quintana}, Elisa V. and {Clarke}, Bruce D. and {Allen}, Christopher and {Li}, Jie and {Wu}, Haley and {Tenenbaum}, Peter and {Verner}, Ekaterina and {Bruhweiler}, Frederick and {Barnes}, Jason and {Prsa}, Andrej},
        title = "{Kepler Planet-Detection Mission: Introduction and First Results}",
      journal = {Science},
         year = 2010,
        month = feb,
       volume = {327},
       number = {5968},
        pages = {977},
          doi = {10.1126/science.1185402},
       adsurl = {https://ui.adsabs.harvard.edu/abs/2010Sci...327..977B}
}

@ARTICLE{2018MNRAS.477.3145J,
       author = {{Jayasinghe}, T. and {Kochanek}, C.~S. and {Stanek}, K.~Z. and {Shappee}, B.~J. and {Holoien}, T.~W.-S. and {Thompson}, Todd A. and {Prieto}, J.~L. and {Dong}, Subo and {Pawlak}, M. and {Shields}, J.~V. and {Pojmanski}, G. and {Otero}, S. and {Britt}, C.~A. and {Will}, D.},
        title = "{The ASAS-SN catalogue of variable stars I: The Serendipitous Survey}",
      journal = {\mnras},
         year = 2018,
        month = jul,
       volume = {477},
       number = {3},
        pages = {3145-3163},
          doi = {10.1093/mnras/sty838},
archivePrefix = {arXiv},
       eprint = {1803.01001},
 primaryClass = {astro-ph.SR},
       adsurl = {https://ui.adsabs.harvard.edu/abs/2018MNRAS.477.3145J}
}

@ARTICLE{2016A&A...595A...1G,
       author = {{Gaia Collaboration} and {Prusti}, T. and {de Bruijne}, J.~H.~J. and {Brown}, A.~G.~A. and {Vallenari}, A. and {Babusiaux}, C. and {Bailer-Jones}, C.~A.~L. and {Bastian}, U. and {Biermann}, M. and {Evans}, D.~W. and {Eyer}, L. and {Jansen}, F. and {Jordi}, C. and {Klioner}, S.~A. and {Lammers}, U. and {Lindegren}, L. and {Luri}, X. and {Mignard}, F. and {Milligan}, D.~J. and {Panem}, C. and {Poinsignon}, V. and {Pourbaix}, D. and {Randich}, S. and {Sarri}, G. and {Sartoretti}, P. and {Siddiqui}, H.~I. and {Soubiran}, C. and {Valette}, V. and {van Leeuwen}, F. and {Walton}, N.~A. and {Aerts}, C. and {Arenou}, F. and {Cropper}, M. and {Drimmel}, R. and {H{\o}g}, E. and {Katz}, D. and {Lattanzi}, M.~G. and {O'Mullane}, W. and {Grebel}, E.~K. and {Holland}, A.~D. and {Huc}, C. and {Passot}, X. and {Bramante}, L. and {Cacciari}, C. and {Casta{\~n}eda}, J. and {Chaoul}, L. and {Cheek}, N. and {De Angeli}, F. and {Fabricius}, C. and {Guerra}, R. and {Hern{\'a}ndez}, J. and {Jean-Antoine-Piccolo}, A. and {Masana}, E. and {Messineo}, R. and {Mowlavi}, N. and {Nienartowicz}, K. and {Ord{\'o}{\~n}ez-Blanco}, D. and {Panuzzo}, P. and {Portell}, J. and {Richards}, P.~J. and {Riello}, M. and {Seabroke}, G.~M. and {Tanga}, P. and {Th{\'e}venin}, F. and {Torra}, J. and {Els}, S.~G. and {Gracia-Abril}, G. and {Comoretto}, G. and {Garcia-Reinaldos}, M. and {Lock}, T. and {Mercier}, E. and {Altmann}, M. and {Andrae}, R. and {Astraatmadja}, T.~L. and {Bellas-Velidis}, I. and {Benson}, K. and {Berthier}, J. and {Blomme}, R. and {Busso}, G. and {Carry}, B. and {Cellino}, A. and {Clementini}, G. and {Cowell}, S. and {Creevey}, O. and {Cuypers}, J. and {Davidson}, M. and {De Ridder}, J. and {de Torres}, A. and {Delchambre}, L. and {Dell'Oro}, A. and {Ducourant}, C. and {Fr{\'e}mat}, Y. and {Garc{\'\i}a-Torres}, M. and {Gosset}, E. and {Halbwachs}, J.-L. and {Hambly}, N.~C. and {Harrison}, D.~L. and {Hauser}, M. and {Hestroffer}, D. and {Hodgkin}, S.~T. and {Huckle}, H.~E. and {Hutton}, A. and {Jasniewicz}, G. and {Jordan}, S. and {Kontizas}, M. and {Korn}, A.~J. and {Lanzafame}, A.~C. and {Manteiga}, M. and {Moitinho}, A. and {Muinonen}, K. and {Osinde}, J. and {Pancino}, E. and {Pauwels}, T. and {Petit}, J.-M. and {Recio-Blanco}, A. and {Robin}, A.~C. and {Sarro}, L.~M. and {Siopis}, C. and {Smith}, M. and {Smith}, K.~W. and {Sozzetti}, A. and {Thuillot}, W. and {van Reeven}, W. and {Viala}, Y. and {Abbas}, U. and {Abreu Aramburu}, A. and {Accart}, S. and {Aguado}, J.~J. and {Allan}, P.~M. and {Allasia}, W. and {Altavilla}, G. and {{\'A}lvarez}, M.~A. and {Alves}, J. and {Anderson}, R.~I. and {Andrei}, A.~H. and {Anglada Varela}, E. and {Antiche}, E. and {Antoja}, T. and {Ant{\'o}n}, S. and {Arcay}, B. and {Atzei}, A. and {Ayache}, L. and {Bach}, N. and {Baker}, S.~G. and {Balaguer-N{\'u}{\~n}ez}, L. and {Barache}, C. and {Barata}, C. and {Barbier}, A. and {Barblan}, F. and {Baroni}, M. and {Barrado y Navascu{\'e}s}, D. and {Barros}, M. and {Barstow}, M.~A. and {Becciani}, U. and {Bellazzini}, M. and {Bellei}, G. and {Bello Garc{\'\i}a}, A. and {Belokurov}, V. and {Bendjoya}, P. and {Berihuete}, A. and {Bianchi}, L. and {Bienaym{\'e}}, O. and {Billebaud}, F. and {Blagorodnova}, N. and {Blanco-Cuaresma}, S. and {Boch}, T. and {Bombrun}, A. and {Borrachero}, R. and {Bouquillon}, S. and {Bourda}, G. and {Bouy}, H. and {Bragaglia}, A. and {Breddels}, M.~A. and {Brouillet}, N. and {Br{\"u}semeister}, T. and {Bucciarelli}, B. and {Budnik}, F. and {Burgess}, P. and {Burgon}, R. and {Burlacu}, A. and {Busonero}, D. and {Buzzi}, R. and {Caffau}, E. and {Cambras}, J. and {Campbell}, H. and {Cancelliere}, R. and {Cantat-Gaudin}, T. and {Carlucci}, T. and {Carrasco}, J.~M. and {Castellani}, M. and {Charlot}, P. and {Charnas}, J. and {Charvet}, P. and {Chassat}, F. and {Chiavassa}, A. and {Clotet}, M. and {Cocozza}, G. and {Collins}, R.~S. and {Collins}, P. and {Costigan}, G.},
        title = "{The Gaia mission}",
      journal = {\aap},
         year = 2016,
        month = nov,
       volume = {595},
          eid = {A1},
        pages = {A1},
          doi = {10.1051/0004-6361/201629272},
archivePrefix = {arXiv},
       eprint = {1609.04153},
 primaryClass = {astro-ph.IM},
       adsurl = {https://ui.adsabs.harvard.edu/abs/2016A&A...595A...1G}
}

@ARTICLE{2019PASP..131a8002B,
       author = {{Bellm}, Eric C. and {Kulkarni}, Shrinivas R. and {Graham}, Matthew J. and {Dekany}, Richard and {Smith}, Roger M. and {Riddle}, Reed and {Masci}, Frank J. and {Helou}, George and {Prince}, Thomas A. and {Adams}, Scott M. and {Barbarino}, C. and {Barlow}, Tom and {Bauer}, James and {Beck}, Ron and {Belicki}, Justin and {Biswas}, Rahul and {Blagorodnova}, Nadejda and {Bodewits}, Dennis and {Bolin}, Bryce and {Brinnel}, Valery and {Brooke}, Tim and {Bue}, Brian and {Bulla}, Mattia and {Burruss}, Rick and {Cenko}, S. Bradley and {Chang}, Chan-Kao and {Connolly}, Andrew and {Coughlin}, Michael and {Cromer}, John and {Cunningham}, Virginia and {De}, Kishalay and {Delacroix}, Alex and {Desai}, Vandana and {Duev}, Dmitry A. and {Eadie}, Gwendolyn and {Farnham}, Tony L. and {Feeney}, Michael and {Feindt}, Ulrich and {Flynn}, David and {Franckowiak}, Anna and {Frederick}, S. and {Fremling}, C. and {Gal-Yam}, Avishay and {Gezari}, Suvi and {Giomi}, Matteo and {Goldstein}, Daniel A. and {Golkhou}, V. Zach and {Goobar}, Ariel and {Groom}, Steven and {Hacopians}, Eugean and {Hale}, David and {Henning}, John and {Ho}, Anna Y.~Q. and {Hover}, David and {Howell}, Justin and {Hung}, Tiara and {Huppenkothen}, Daniela and {Imel}, David and {Ip}, Wing-Huen and {Ivezi{\'c}}, {\v{Z}}eljko and {Jackson}, Edward and {Jones}, Lynne and {Juric}, Mario and {Kasliwal}, Mansi M. and {Kaspi}, S. and {Kaye}, Stephen and {Kelley}, Michael S.~P. and {Kowalski}, Marek and {Kramer}, Emily and {Kupfer}, Thomas and {Landry}, Walter and {Laher}, Russ R. and {Lee}, Chien-De and {Lin}, Hsing Wen and {Lin}, Zhong-Yi and {Lunnan}, Ragnhild and {Giomi}, Matteo and {Mahabal}, Ashish and {Mao}, Peter and {Miller}, Adam A. and {Monkewitz}, Serge and {Murphy}, Patrick and {Ngeow}, Chow-Choong and {Nordin}, Jakob and {Nugent}, Peter and {Ofek}, Eran and {Patterson}, Maria T. and {Penprase}, Bryan and {Porter}, Michael and {Rauch}, Ludwig and {Rebbapragada}, Umaa and {Reiley}, Dan and {Rigault}, Mickael and {Rodriguez}, Hector and {van Roestel}, Jan and {Rusholme}, Ben and {van Santen}, Jakob and {Schulze}, S. and {Shupe}, David L. and {Singer}, Leo P. and {Soumagnac}, Maayane T. and {Stein}, Robert and {Surace}, Jason and {Sollerman}, Jesper and {Szkody}, Paula and {Taddia}, F. and {Terek}, Scott and {Van Sistine}, Angela and {van Velzen}, Sjoert and {Vestrand}, W. Thomas and {Walters}, Richard and {Ward}, Charlotte and {Ye}, Quan-Zhi and {Yu}, Po-Chieh and {Yan}, Lin and {Zolkower}, Jeffry},
        title = "{The Zwicky Transient Facility: System Overview, Performance, and First Results}",
      journal = {\pasp},
         year = 2019,
        month = jan,
       volume = {131},
       number = {995},
        pages = {018002},
          doi = {10.1088/1538-3873/aaecbe},
archivePrefix = {arXiv},
       eprint = {1902.01932},
 primaryClass = {astro-ph.IM},
       adsurl = {https://ui.adsabs.harvard.edu/abs/2019PASP..131a8002B}
}

@ARTICLE{2019PASP..131a8003M,
       author = {{Masci}, Frank J. and {Laher}, Russ R. and {Rusholme}, Ben and {Shupe}, David L. and {Groom}, Steven and {Surace}, Jason and {Jackson}, Edward and {Monkewitz}, Serge and {Beck}, Ron and {Flynn}, David and {Terek}, Scott and {Landry}, Walter and {Hacopians}, Eugean and {Desai}, Vandana and {Howell}, Justin and {Brooke}, Tim and {Imel}, David and {Wachter}, Stefanie and {Ye}, Quan-Zhi and {Lin}, Hsing-Wen and {Cenko}, S. Bradley and {Cunningham}, Virginia and {Rebbapragada}, Umaa and {Bue}, Brian and {Miller}, Adam A. and {Mahabal}, Ashish and {Bellm}, Eric C. and {Patterson}, Maria T. and {Juri{\'c}}, Mario and {Golkhou}, V. Zach and {Ofek}, Eran O. and {Walters}, Richard and {Graham}, Matthew and {Kasliwal}, Mansi M. and {Dekany}, Richard G. and {Kupfer}, Thomas and {Burdge}, Kevin and {Cannella}, Christopher B. and {Barlow}, Tom and {Van Sistine}, Angela and {Giomi}, Matteo and {Fremling}, Christoffer and {Blagorodnova}, Nadejda and {Levitan}, David and {Riddle}, Reed and {Smith}, Roger M. and {Helou}, George and {Prince}, Thomas A. and {Kulkarni}, Shrinivas R.},
        title = "{The Zwicky Transient Facility: Data Processing, Products, and Archive}",
      journal = {\pasp},
         year = 2019,
        month = jan,
       volume = {131},
       number = {995},
        pages = {018003},
          doi = {10.1088/1538-3873/aae8ac},
archivePrefix = {arXiv},
       eprint = {1902.01872},
 primaryClass = {astro-ph.IM},
       adsurl = {https://ui.adsabs.harvard.edu/abs/2019PASP..131a8003M}
}

@ARTICLE{2021A&C....3600488C,
       author = {{{\v{C}}okina}, M. and {Maslej-Kre{\v{s}}{\v{n}}{\'a}kov{\'a}}, V. and {Butka}, P. and {Parimucha}, {\v{S}}.},
        title = "{Automatic classification of eclipsing binary stars using deep learning methods}",
      journal = {Astronomy and Computing},
         year = 2021,
        month = jul,
       volume = {36},
          eid = {100488},
        pages = {100488},
          doi = {10.1016/j.ascom.2021.100488},
archivePrefix = {arXiv},
       eprint = {2108.01640},
 primaryClass = {astro-ph.SR},
       adsurl = {https://ui.adsabs.harvard.edu/abs/2021A&C....3600488C}
}

@ARTICLE{2010NewA...15..433M,
       author = {{Minniti}, D. and {Lucas}, P.~W. and {Emerson}, J.~P. and {Saito}, R.~K. and {Hempel}, M. and {Pietrukowicz}, P. and {Ahumada}, A.~V. and {Alonso}, M.~V. and {Alonso-Garcia}, J. and {Arias}, J.~I. and {Bandyopadhyay}, R.~M. and {Barb{\'a}}, R.~H. and {Barbuy}, B. and {Bedin}, L.~R. and {Bica}, E. and {Borissova}, J. and {Bronfman}, L. and {Carraro}, G. and {Catelan}, M. and {Clari{\'a}}, J.~J. and {Cross}, N. and {de Grijs}, R. and {D{\'e}k{\'a}ny}, I. and {Drew}, J.~E. and {Fari{\~n}a}, C. and {Feinstein}, C. and {Fern{\'a}ndez Laj{\'u}s}, E. and {Gamen}, R.~C. and {Geisler}, D. and {Gieren}, W. and {Goldman}, B. and {Gonzalez}, O.~A. and {Gunthardt}, G. and {Gurovich}, S. and {Hambly}, N.~C. and {Irwin}, M.~J. and {Ivanov}, V.~D. and {Jord{\'a}n}, A. and {Kerins}, E. and {Kinemuchi}, K. and {Kurtev}, R. and {L{\'o}pez-Corredoira}, M. and {Maccarone}, T. and {Masetti}, N. and {Merlo}, D. and {Messineo}, M. and {Mirabel}, I.~F. and {Monaco}, L. and {Morelli}, L. and {Padilla}, N. and {Palma}, T. and {Parisi}, M.~C. and {Pignata}, G. and {Rejkuba}, M. and {Roman-Lopes}, A. and {Sale}, S.~E. and {Schreiber}, M.~R. and {Schr{\"o}der}, A.~C. and {Smith}, M. and {Sodr{\'e}}, Jr., L. and {Soto}, M. and {Tamura}, M. and {Tappert}, C. and {Thompson}, M.~A. and {Toledo}, I. and {Zoccali}, M. and {Pietrzynski}, G.},
        title = "{VISTA Variables in the Via Lactea (VVV): The public ESO near-IR variability survey of the Milky Way}",
      journal = {\na},
         year = 2010,
        month = jul,
       volume = {15},
       number = {5},
        pages = {433-443},
          doi = {10.1016/j.newast.2009.12.002},
archivePrefix = {arXiv},
       eprint = {0912.1056},
 primaryClass = {astro-ph.GA},
       adsurl = {https://ui.adsabs.harvard.edu/abs/2010NewA...15..433M}
}

@ARTICLE{2025ApJS..277...51L,
       author = {{Li}, Kai and {Wang}, Li-Heng},
        title = "{Physical Parameters of 12,201 ASAS-SN Contact Binaries Determined by a Neural Network}",
      journal = {\apjs},
         year = 2025,
        month = apr,
       volume = {277},
       number = {2},
          eid = {51},
        pages = {51},
          doi = {10.3847/1538-4365/adba63},
archivePrefix = {arXiv},
       eprint = {2502.16206},
 primaryClass = {astro-ph.SR},
       adsurl = {https://ui.adsabs.harvard.edu/abs/2025ApJS..277...51L}
}

@ARTICLE{1965ApJS...11..216L,
       author = {{Lafler}, J. and {Kinman}, T.~D.},
        title = "{An RR Lyrae Star Survey with Ihe Lick 20-INCH Astrograph II. The Calculation of RR Lyrae Periods by Electronic Computer.}",
      journal = {\apjs},
         year = 1965,
        month = jun,
       volume = {11},
        pages = {216},
          doi = {10.1086/190116},
       adsurl = {https://ui.adsabs.harvard.edu/abs/1965ApJS...11..216L}
}

@ARTICLE{2002A&A...386..763C,
       author = {{Clarke}, D.},
        title = "{String/Rope length methods using the Lafler-Kinman statistic}",
      journal = {\aap},
         year = 2002,
        month = may,
       volume = {386},
        pages = {763-774},
          doi = {10.1051/0004-6361:20020258},
       adsurl = {https://ui.adsabs.harvard.edu/abs/2002A&A...386..763C}
}

@ARTICLE{2014PASP..126..398H,
       author = {{Howell}, Steve B. and {Sobeck}, Charlie and {Haas}, Michael and {Still}, Martin and {Barclay}, Thomas and {Mullally}, Fergal and {Troeltzsch}, John and {Aigrain}, Suzanne and {Bryson}, Stephen T. and {Caldwell}, Doug and {Chaplin}, William J. and {Cochran}, William D. and {Huber}, Daniel and {Marcy}, Geoffrey W. and {Miglio}, Andrea and {Najita}, Joan R. and {Smith}, Marcie and {Twicken}, J.~D. and {Fortney}, Jonathan J.},
        title = "{The K2 Mission: Characterization and Early Results}",
      journal = {\pasp},
         year = 2014,
        month = apr,
       volume = {126},
       number = {938},
        pages = {398},
          doi = {10.1086/676406},
archivePrefix = {arXiv},
       eprint = {1402.5163},
 primaryClass = {astro-ph.IM},
       adsurl = {https://ui.adsabs.harvard.edu/abs/2014PASP..126..398H}
}

@ARTICLE{1991A&ARv...3...91A,
       author = {{Andersen}, J.},
        title = "{Accurate masses and radii of normal stars}",
      journal = {\aapr},
         year = 1991,
        month = jan,
       volume = {3},
       number = {2},
        pages = {91-126},
          doi = {10.1007/BF00873538},
       adsurl = {https://ui.adsabs.harvard.edu/abs/1991A&ARv...3...91A}
}

@ARTICLE{1986Natur.323..533R,
       author = {{Rumelhart}, David E. and {Hinton}, Geoffrey E. and {Williams}, Ronald J.},
        title = "{Learning representations by back-propagating errors}",
      journal = {\nat},
         year = 1986,
        month = oct,
       volume = {323},
       number = {6088},
        pages = {533-536},
          doi = {10.1038/323533a0},
       adsurl = {https://ui.adsabs.harvard.edu/abs/1986Natur.323..533R}
}

@software{2018ascl.soft12013L,
       author = {{Lightkurve Collaboration} and {Cardoso}, Jos{\'e} Vin{\'\i}cius de Miranda and {Hedges}, Christina and {Gully-Santiago}, Michael and {Saunders}, Nicholas and {Cody}, Ann Marie and {Barclay}, Thomas and {Hall}, Oliver and {Sagear}, Sheila and {Turtelboom}, Emma and {Zhang}, Johnny and {Tzanidakis}, Andy and {Mighell}, Ken and {Coughlin}, Jeff and {Bell}, Keaton and {Berta-Thompson}, Zach and {Williams}, Peter and {Dotson}, Jessie and {Barentsen}, Geert},
        title = "{Lightkurve: Kepler and TESS time series analysis in Python}",
 howpublished = {Astrophysics Source Code Library, record ascl:1812.013},
         year = 2018,
        month = dec,
          eid = {ascl:1812.013},
archivePrefix = {ascl},
       eprint = {1812.013},
       adsurl = {https://ui.adsabs.harvard.edu/abs/2018ascl.soft12013L}
}

@inproceedings{10.5555/3104322.3104425,
author = {Nair, Vinod and Hinton, Geoffrey E.},
title = {Rectified linear units improve restricted boltzmann machines},
year = {2010},
isbn = {9781605589077},
publisher = {Omnipress},
address = {Madison, WI, USA},
booktitle = {Proceedings of the 27th International Conference on International Conference on Machine Learning},
pages = {807–814},
numpages = {8},
location = {Haifa, Israel},
series = {ICML'10}
}

@article{10.5555/2627435.2670313,
author = {Srivastava, Nitish and Hinton, Geoffrey and Krizhevsky, Alex and Sutskever, Ilya and Salakhutdinov, Ruslan},
title = {Dropout: a simple way to prevent neural networks from overfitting},
year = {2014},
issue_date = {January 2014},
publisher = {JMLR.org},
volume = {15},
number = {1},
issn = {1532-4435},
journal = {J. Mach. Learn. Res.},
month = jan,
pages = {1929–1958},
numpages = {30}
}

@ARTICLE{2014arXiv1412.6980K,
       author = {{Kingma}, Diederik P. and {Ba}, Jimmy},
        title = "{Adam: A Method for Stochastic Optimization}",
      journal = {arXiv e-prints},
         year = 2014,
        month = dec,
          eid = {arXiv:1412.6980},
        pages = {arXiv:1412.6980},
          doi = {10.48550/arXiv.1412.6980},
archivePrefix = {arXiv},
       eprint = {1412.6980},
 primaryClass = {cs.LG},
       adsurl = {https://ui.adsabs.harvard.edu/abs/2014arXiv1412.6980K}
}

@ARTICLE{2006Sci...313..504H,
       author = {{Hinton}, G.~E. and {Salakhutdinov}, R.~R.},
        title = "{Reducing the Dimensionality of Data with Neural Networks}",
      journal = {Science},
         year = 2006,
        month = jul,
       volume = {313},
       number = {5786},
        pages = {504-507},
          doi = {10.1126/science.1127647},
       adsurl = {https://ui.adsabs.harvard.edu/abs/2006Sci...313..504H}
}

@article{10.1162/089976601750264965,
author = {Sch\"{o}lkopf, Bernhard and Platt, John C. and Shawe-Taylor, John C. and Smola, Alex J. and Williamson, Robert C.},
title = {Estimating the Support of a High-Dimensional Distribution},
year = {2001},
issue_date = {July 2001},
publisher = {MIT Press},
address = {Cambridge, MA, USA},
volume = {13},
number = {7},
issn = {0899-7667},
url = {https://doi.org/10.1162/089976601750264965},
doi = {10.1162/089976601750264965},
journal = {Neural Comput.},
month = jul,
pages = {1443–1471},
numpages = {29}
}

@article{broomhead1988multivariable,
  title={Multivariable functional interpolation and adaptive networks},
  author={Broomhead, David S and Lowe, David},
  journal={Complex Systems},
  volume={2},
  number={3},
  pages={321--355},
  year={1988}
}

@ARTICLE{1992AJ....103..960R,
       author = {{Rucinski}, S.~M.},
        title = "{Can Full Convection Explain the Observed Short-Period Limit of the W UMa-Type Binaries?}",
      journal = {\aj},
         year = 1992,
        month = mar,
       volume = {103},
        pages = {960},
          doi = {10.1086/116118},
       adsurl = {https://ui.adsabs.harvard.edu/abs/1992AJ....103..960R}
}

@ARTICLE{2012MNRAS.421.2769J,
       author = {{Jiang}, Dengkai and {Han}, Zhanwen and {Ge}, Hongwei and {Yang}, Liheng and {Li}, Lifang},
        title = "{The short-period limit of contact binaries}",
      journal = {\mnras},
         year = 2012,
        month = apr,
       volume = {421},
       number = {4},
        pages = {2769-2773},
          doi = {10.1111/j.1365-2966.2011.20323.x},
archivePrefix = {arXiv},
       eprint = {1112.0466},
 primaryClass = {astro-ph.SR},
       adsurl = {https://ui.adsabs.harvard.edu/abs/2012MNRAS.421.2769J}
}

@ARTICLE{2015AJ....150..117Q,
       author = {{Qian}, S.~B. and {Zhang}, B. and {Soonthornthum}, B. and {He}, J.~J. and {Rattanasoon}, S. and {Aukkaravittayapun}, S. and {Liu}, L. and {Zhu}, L.~Y. and {Zhao}, E.~G. and {Zhou}, X. and {Thawicharat}, S.},
        title = "{SuperWASP J015100.23-100524.2: A Spotted Shallow-contact Binary Below the Period Limit}",
      journal = {\aj},
         year = 2015,
        month = oct,
       volume = {150},
       number = {4},
          eid = {117},
        pages = {117},
          doi = {10.1088/0004-6256/150/4/117},
       adsurl = {https://ui.adsabs.harvard.edu/abs/2015AJ....150..117Q}
}

@ARTICLE{2020AJ....159..189L,
       author = {{Li}, Kai and {Kim}, Chun-Hwey and {Xia}, Qi-Qi and {Michel}, Raul and {Hu}, Shao-Ming and {Gao}, Xing and {Guo}, Di-Fu and {Chen}, Xu},
        title = "{The First Light Curve Modeling and Orbital Period Change Investigation of Nine Contact Binaries around the Short-period Cutoff}",
      journal = {\aj},
         year = 2020,
        month = may,
       volume = {159},
       number = {5},
          eid = {189},
        pages = {189},
          doi = {10.3847/1538-3881/ab7cda},
archivePrefix = {arXiv},
       eprint = {2003.02377},
 primaryClass = {astro-ph.SR},
       adsurl = {https://ui.adsabs.harvard.edu/abs/2020AJ....159..189L}
}

@book{Kallrath2009Eclipsing,
  author    = {Kallrath, Josef and Milone, Eugene F.},
  title     = {Eclipsing Binary Stars: Modeling and Analysis},
  edition   = {2},
  series    = {Astronomy and Astrophysics Library},
  publisher = {Springer},
  year      = {2009},
  address   = {New York},
  isbn      = {978-1-4419-0698-4},
  doi       = {10.1007/978-1-4419-0699-1},
}

@ARTICLE{2019ApJ...873..111I,
       author = {{Ivezi{\'c}}, {\v{Z}}eljko and {Kahn}, Steven M. and {Tyson}, J. Anthony and {Abel}, Bob and {Acosta}, Emily and {Allsman}, Robyn and {Alonso}, David and {AlSayyad}, Yusra and {Anderson}, Scott F. and {Andrew}, John and {Angel}, James Roger P. and {Angeli}, George Z. and {Ansari}, Reza and {Antilogus}, Pierre and {Araujo}, Constanza and {Armstrong}, Robert and {Arndt}, Kirk T. and {Astier}, Pierre and {Aubourg}, {\'E}ric and {Auza}, Nicole and {Axelrod}, Tim S. and {Bard}, Deborah J. and {Barr}, Jeff D. and {Barrau}, Aurelian and {Bartlett}, James G. and {Bauer}, Amanda E. and {Bauman}, Brian J. and {Baumont}, Sylvain and {Bechtol}, Ellen and {Bechtol}, Keith and {Becker}, Andrew C. and {Becla}, Jacek and {Beldica}, Cristina and {Bellavia}, Steve and {Bianco}, Federica B. and {Biswas}, Rahul and {Blanc}, Guillaume and {Blazek}, Jonathan and {Blandford}, Roger D. and {Bloom}, Josh S. and {Bogart}, Joanne and {Bond}, Tim W. and {Booth}, Michael T. and {Borgland}, Anders W. and {Borne}, Kirk and {Bosch}, James F. and {Boutigny}, Dominique and {Brackett}, Craig A. and {Bradshaw}, Andrew and {Brandt}, William Nielsen and {Brown}, Michael E. and {Bullock}, James S. and {Burchat}, Patricia and {Burke}, David L. and {Cagnoli}, Gianpietro and {Calabrese}, Daniel and {Callahan}, Shawn and {Callen}, Alice L. and {Carlin}, Jeffrey L. and {Carlson}, Erin L. and {Chandrasekharan}, Srinivasan and {Charles-Emerson}, Glenaver and {Chesley}, Steve and {Cheu}, Elliott C. and {Chiang}, Hsin-Fang and {Chiang}, James and {Chirino}, Carol and {Chow}, Derek and {Ciardi}, David R. and {Claver}, Charles F. and {Cohen-Tanugi}, Johann and {Cockrum}, Joseph J. and {Coles}, Rebecca and {Connolly}, Andrew J. and {Cook}, Kem H. and {Cooray}, Asantha and {Covey}, Kevin R. and {Cribbs}, Chris and {Cui}, Wei and {Cutri}, Roc and {Daly}, Philip N. and {Daniel}, Scott F. and {Daruich}, Felipe and {Daubard}, Guillaume and {Daues}, Greg and {Dawson}, William and {Delgado}, Francisco and {Dellapenna}, Alfred and {de Peyster}, Robert and {de Val-Borro}, Miguel and {Digel}, Seth W. and {Doherty}, Peter and {Dubois}, Richard and {Dubois-Felsmann}, Gregory P. and {Durech}, Josef and {Economou}, Frossie and {Eifler}, Tim and {Eracleous}, Michael and {Emmons}, Benjamin L. and {Fausti Neto}, Angelo and {Ferguson}, Henry and {Figueroa}, Enrique and {Fisher-Levine}, Merlin and {Focke}, Warren and {Foss}, Michael D. and {Frank}, James and {Freemon}, Michael D. and {Gangler}, Emmanuel and {Gawiser}, Eric and {Geary}, John C. and {Gee}, Perry and {Geha}, Marla and {Gessner}, Charles J.~B. and {Gibson}, Robert R. and {Gilmore}, D. Kirk and {Glanzman}, Thomas and {Glick}, William and {Goldina}, Tatiana and {Goldstein}, Daniel A. and {Goodenow}, Iain and {Graham}, Melissa L. and {Gressler}, William J. and {Gris}, Philippe and {Guy}, Leanne P. and {Guyonnet}, Augustin and {Haller}, Gunther and {Harris}, Ron and {Hascall}, Patrick A. and {Haupt}, Justine and {Hernandez}, Fabio and {Herrmann}, Sven and {Hileman}, Edward and {Hoblitt}, Joshua and {Hodgson}, John A. and {Hogan}, Craig and {Howard}, James D. and {Huang}, Dajun and {Huffer}, Michael E. and {Ingraham}, Patrick and {Innes}, Walter R. and {Jacoby}, Suzanne H. and {Jain}, Bhuvnesh and {Jammes}, Fabrice and {Jee}, M. James and {Jenness}, Tim and {Jernigan}, Garrett and {Jevremovi{\'c}}, Darko and {Johns}, Kenneth and {Johnson}, Anthony S. and {Johnson}, Margaret W.~G. and {Jones}, R. Lynne and {Juramy-Gilles}, Claire and {Juri{\'c}}, Mario and {Kalirai}, Jason S. and {Kallivayalil}, Nitya J. and {Kalmbach}, Bryce and {Kantor}, Jeffrey P. and {Karst}, Pierre and {Kasliwal}, Mansi M. and {Kelly}, Heather and {Kessler}, Richard and {Kinnison}, Veronica and {Kirkby}, David and {Knox}, Lloyd and {Kotov}, Ivan V. and {Krabbendam}, Victor L. and {Krughoff}, K. Simon and {Kub{\'a}nek}, Petr and {Kuczewski}, John and {Kulkarni}, Shri and {Ku}, John and {Kurita}, Nadine R. and {Lage}, Craig S. and {Lambert}, Ron and {Lange}, Travis and {Langton}, J. Brian and {Le Guillou}, Laurent and {Levine}, Deborah and {Liang}, Ming and {Lim}, Kian-Tat and {Lintott}, Chris J. and {Long}, Kevin E. and {Lopez}, Margaux and {Lotz}, Paul J. and {Lupton}, Robert H. and {Lust}, Nate B. and {MacArthur}, Lauren A. and {Mahabal}, Ashish and {Mandelbaum}, Rachel and {Markiewicz}, Thomas W. and {Marsh}, Darren S. and {Marshall}, Philip J. and {Marshall}, Stuart and {May}, Morgan and {McKercher}, Robert and {McQueen}, Michelle and {Meyers}, Joshua and {Migliore}, Myriam and {Miller}, Michelle and {Mills}, David J.},
        title = "{LSST: From Science Drivers to Reference Design and Anticipated Data Products}",
      journal = {\apj},
         year = 2019,
        month = mar,
       volume = {873},
       number = {2},
          eid = {111},
        pages = {111},
          doi = {10.3847/1538-4357/ab042c},
archivePrefix = {arXiv},
       eprint = {0805.2366},
 primaryClass = {astro-ph},
       adsurl = {https://ui.adsabs.harvard.edu/abs/2019ApJ...873..111I}
}

@book{Goodfellow2016Deep,
  title     = {Deep Learning},
  author    = {Goodfellow, Ian and Bengio, Yoshua and Courville, Aaron},
  year      = {2016},
  publisher = {MIT Press},
  series    = {Adaptive Computation and Machine Learning Series},
  isbn      = {978-0262035613},
  url       = {http://www.deeplearningbook.org/}
}
\bibliographystyle{aasjournalv7}

\end{document}